\documentclass[reprint,groupedaddress,superscriptaddress,amsmath,amssymb,prl]{revtex4-2}

\usepackage[utf8]{inputenc}
\usepackage{graphicx}
\usepackage{dcolumn}
\usepackage{bm}
\usepackage[%
    colorlinks=true,
    urlcolor=magenta,
    linkcolor=red,
    citecolor=cyan
]{hyperref}
\usepackage[mathlines]{lineno}
\usepackage{siunitx}
\usepackage{amsmath}
\usepackage{physics}
\usepackage{multirow}
\usepackage{kotex}
\usepackage[nodisplayskipstretch]{setspace}

\usepackage[most]{tcolorbox}
\usepackage{xcolor}
\usepackage{soul}

\usepackage{bm}
\newcommand{\circled}[2][]{%
  \tikz[baseline=(char.base)]{%
    \node[shape = circle, draw, inner sep = 0.1pt]
    (char) {\phantom{\ifblank{#1}{#2}{#1}}};%
    \node at (char.center) {\makebox[0pt][c]{#2}};}}
\robustify{\circled}

\begin{document}
\title{Bound States-to-Bands in the Continuum in Cylindrical Granular Crystals}

 \author{Yeongtae Jang}
     \affiliation{%
     Department of Mechanical Engineering, Pohang University of Science and Technology (POSTECH), Pohang 37673, Republic of Korea}
     \email{jrsho@postech.ac.kr}
 \author{Seokwoo Kim}
     \affiliation{%
     Department of Mechanical Engineering, Pohang University of Science and Technology (POSTECH), Pohang 37673, Republic of Korea}
 \author{Dongwoo Lee}
     \affiliation{%
     Department of Mechanical Engineering, Pohang University of Science and Technology (POSTECH), Pohang 37673, Republic of Korea}
  \author{Eunho Kim}\email{eunhokim@jbnu.ac.kr}
      \affiliation{%
        Division of Mechanical System Engineering, Jeonbuk National University, Jeonju, 54896, Republic of Korea}
      \email{eunhokim@jbnu.ac.kr}
      \affiliation{%
     Department of JBNU-KIST Industry-Academia Convergence Research, Jeonju, 54896, Republic of Korea}
 \author{Junsuk Rho}\email{jrsho@postech.ac.kr}
     \affiliation{%
     Department of Mechanical Engineering, Pohang University of Science and Technology (POSTECH), Pohang 37673, Republic of Korea}
     \affiliation{%
     Department of Chemical Engineering, Pohang University of Science and Technology (POSTECH), Pohang 37673, Republic of Korea}
    \affiliation{%
    Department of Electrical Engineering, Pohang University of Science and Technology (POSTECH), Pohang 37673, Republic of Korea}
     \affiliation{%
     POSCO-POSTECH-RIST Convergence Research Center for Flat Optics and Metaphotonics, Pohang 37673, Korea}

\date{\today}
\begin{abstract}
We theoretically investigate and experimentally demonstrate that genuine bound states in the continuum (BICs)---polarization-protected BICs---can be completely localized within finite-size solid resonators.
This bound mode is realized in a highly tunable mechanical system made of cylindrical granular crystals, where tunning the contact boundaries enables the \textit{in situ} transition from the BICs to quasi-BICs in a controllable manner.
Since a single-particle resonator can support BICs itself, these bound states can extend to form bound bands within periodic structures composed of such resonators.
We experimentally demonstrate the emergence of a quasi-bound (flat) band in a finite chain with broken resonator symmetry, using a laser Doppler vibrometer.
Remarkably, we show that all cylindrical resonators within the entire chain exhibit high-Q and dispersionless resonance.
\end{abstract}

\maketitle
Bound states in the continuum (BICs) are counterintuitive localized states that coexist with a continuous spectrum of extended states~\cite{Hsu2016}. 
In 1929, von Neumann and Wigner first predicted the concept of BICs in quantum mechanics~\cite{Neumann_1929}.
Shortly thereafter, the concept of BICs was extended to wave functions in general and realized in systems with special local potentials~\cite{Stillinger1975}.
A key attraction of the BICs is their zero radiative losses, which lead to a diverging radiative quality factor (Q-factor).
This feature has naturally led to various dynamics domains, including quantum dots~\cite{Rotter2005,Moiseyev2009}, photonics~\cite{Marinica2008,Plotnik2011,Lee2012,Schiattarella2024}, hydrodynamics~\cite{Ursell1951,Callan1991,COBELLI2011}, and mechanical systems~\cite{Shipman2010,Huang2021,Lee2023}.
Moreover, the strong localization and adjustable high Q-factor have propelled advancements in applications such as lasing and sensing~\cite{Kodigala2017,Koshelev2020,Huang2020}.
\vspace{1mm}
\\
\indent
Most previous theoretical and experimental studies on the BICs have been investigated on structures that are uniform or periodic in one or more directions (e.g., $x$~and $y$), with the BICs localized in other directions (e.g., $z$).
Due to translational symmetry, the wave vector $\mathbf{k}$=$(k_x, k_y)$ is conserved and enables bound states to remain confined and non-radiative in the $z$-direction. 
Thus, BICs are typically observed at isolated wave vectors.
Such restricted structures are commonly used because the \textit{nonexistence theorem}~\cite{Hsu2016} prohibits genuine BICs in compact resonators for single-particle-like systems, making uniform or periodic configurations essential for realizing BICs.
Nevertheless, there are special exceptions.
For instance, in optics, if the effective material has $\epsilon$=0 and $\mu$=0 (epsilon-near-zero)~\cite{Silveirinha2014,Lannebre2015}, it can act as infinite potential barriers that spatially decouple the internal resonances from the external radiation continuum.
In this regard, a theoretical study using composite spherical particles has been proposed in Ref.~\cite{Monticone2014}; however, the experimental realization of the BICs within a single particle remains elusive due to high losses $\text{Im}(\epsilon)$.
\vspace{1mm}
\\
\indent
More recently, another clever approach for realizing genuine BICs in compact resonators has been proposed in continuum mechanics, referred to as \textit{polarization-protected BICs}~\cite{Deriy2022}.
The solid acoustic resonators exhibit a hybridization of longitudinal and transverse polarizations, meaning that they are coupled to the radiation continuum. 
By designing specially shaped solid resonators with internal displacement fields orthogonal to the wave vector at the resonator \textit{boundary}, one can achieve complete decoupling from the radiation continuum.
Despite progress, the experimental demonstration remains incomplete due to the challenge of precisely controlling the coupling strength of the wave vector at the resonator boundary. 
Ensuring orthogonality to the radiation continuum across all boundaries demands a highly precise design.
Furthermore, the necessity for precise actuation and measurement adds to the complexity of this challenge.
A design capable of addressing this issue would deepen our understanding of genuine BICs and unlock the potential of high-Q resonators for various applications.
\vspace{1mm}
\\
\indent
In this Letter, we demonstrate the realization of polarization-protected BICs using granular systems made of cylindrical particles interacting through contact~\cite{johnson1987contact,Nesterenko_2001}.
This system is highly tunable~\cite{Porter2015,Kim2015,Chong2017}, with the coupling strength between the wave vector at the resonator's boundaries and internal local resonance modes adjustable by modifying the contact shift.
In this way, we achieve BICs in a finite-size resonator, where the resonator supports purely rotational motion with tangential displacement at the contact boundary.
Our theoretical model indicates that these BICs exhibit \textit{mass-near-zero}.
What's more, we propose the concept of bound bands in the continuum (BBICs) driven by finite-size resonators under the periodicity, rather than a singular state from BICs.
We experimentally observe BBICs as quasi-flat bands in a periodic BIC resonator chain with broken symmetry, demonstrating high-Q and dispersionless resonance throughout all resonators.
\begin{figure}
\includegraphics [width=0.49\textwidth]{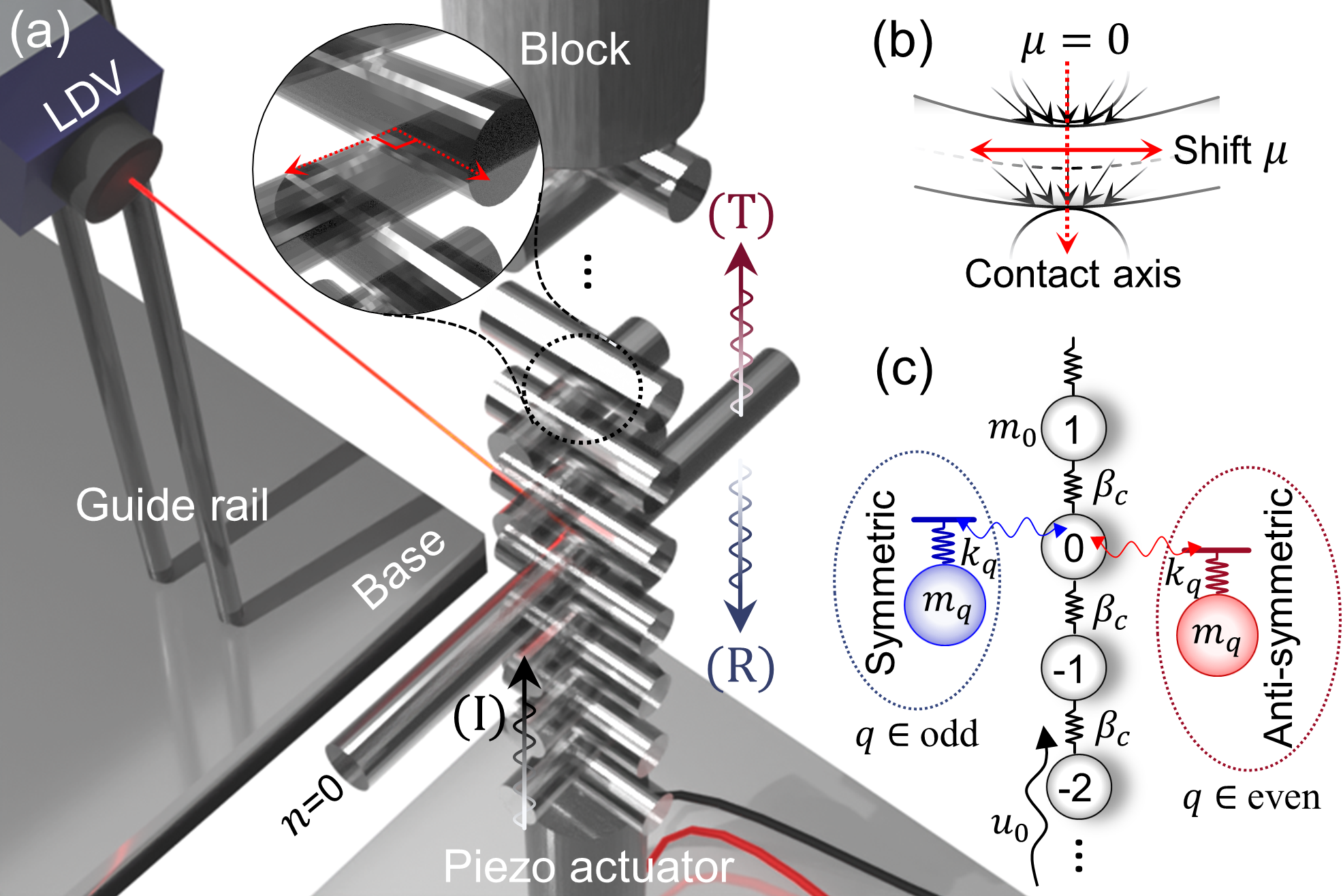}
\caption{Schematic of (a) the experimental setup composed of cylindrical elements used to demonstrate polarization-protected BICs, (b) tuning of the coupling between the resonator displacement and the wave vector at the contact boundary through resonator shifts, and (c) the theoretical discrete element model.}
\label{fig1}
\end{figure}
\vspace{2mm}
\\
\indent
\textit{Experimental and Theoretical Setup.}---
The experimental setup begins with a BIC defect, where a single long cylinder is embedded as a perturbation within a chain of short cylinders [Fig.~\ref{fig1}(a)].
The adoption of finite chains rather than single resonators is motivated by two factors: (i) experimentally, accurately measuring the Q-factor of a single resonator is challenging due to the boundary conditions effect in the setup, and (ii) by utilizing elastic wave propagation along the contact axis within the finite chain, we can define the continuous spectrum of the granular medium.
This, in turn, theoretically allows for a straightforward calculation of the scattering processes between plane waves and BIC defects.
\\
\indent
Initially, we align the center of mass (CoM) of each cylinder to contact neighboring particles, while maintaining orthogonal angles between them. 
All cylinders are made of fused quartz: the host cylinder is 30 mm in length, and the defect cylinder is 90 mm in length, both with an identical diameter of 5 mm.
The long cylindrical element characterizes bending vibrational modes in low-frequency regimes, and its mode deformations are strongly coupled with propagating waves.
We tune the strength of the “mode coupling” between the resonator's vibrational mode displacement and the wave vector at the contact boundary through \textit{in situ} shift of the BIC defect.
This change in coupling occurs because the vibrational mode shape of the resonator affects how the wave interacts with the contact boundary condition.
The shift value for the CoM of the cylinder is $\mu$=0, while one end has $\mu$=1 [Fig.~\ref{fig1}(b)].
To restrict the system dynamics within the linear regime, a free weight (13~\text{N}) is placed on top for pre-compression. 
We excite the first cylinder at the bottom using a piezo-actuator, and the localized vibration at the boundary of the BIC defect cylinder is detected using laser Doppler vibrometry.
\\
\indent
The dynamics of this cylinder chain are analyzed using a discrete element model (DEM) [see~Fig.~\ref{fig1}(c)].
We represent the host cylinders as lumped masses, while the defect BIC cylinder is modeled as a lumped mass coupled with multiple harmonic oscillators, indicating resonance mode coupling.
This mode coupling manifests in the bending modes of the cylindrical elements, which are classified into symmetric and anti-symmetric modes based on deformation symmetry with respect to CoM (see the Supplemental Material~\cite{supp}).
The coefficients for the harmonic oscillators ($m_q$,~$k_q$, where $q$ is the mode number) are determined using physics-informed discrete element modeling~\cite{Jang_2024}, an analytical approach for mapping a continuum beam to a discretized unit cell.
The interactions between neighboring cylinders at their contact are represented by springs ($\beta_c$) that follow Hertz's law~\cite{johnson1987contact}.
If the forces from external perturbations are much smaller than the pre-compression, we can assume linear dynamics, as is the case here; $\beta_c$ is the linearized contact stiffness.
Using this DEM, we analyze the scattering problem and evaluate the Q-factor of the system's steady-state response, comparing it with experimental data (see the Supplemental Material~\cite{supp} for relevant calculations of wave dynamics).
\section{Results and discussions}
\begin{figure}
\includegraphics [width=0.49\textwidth]{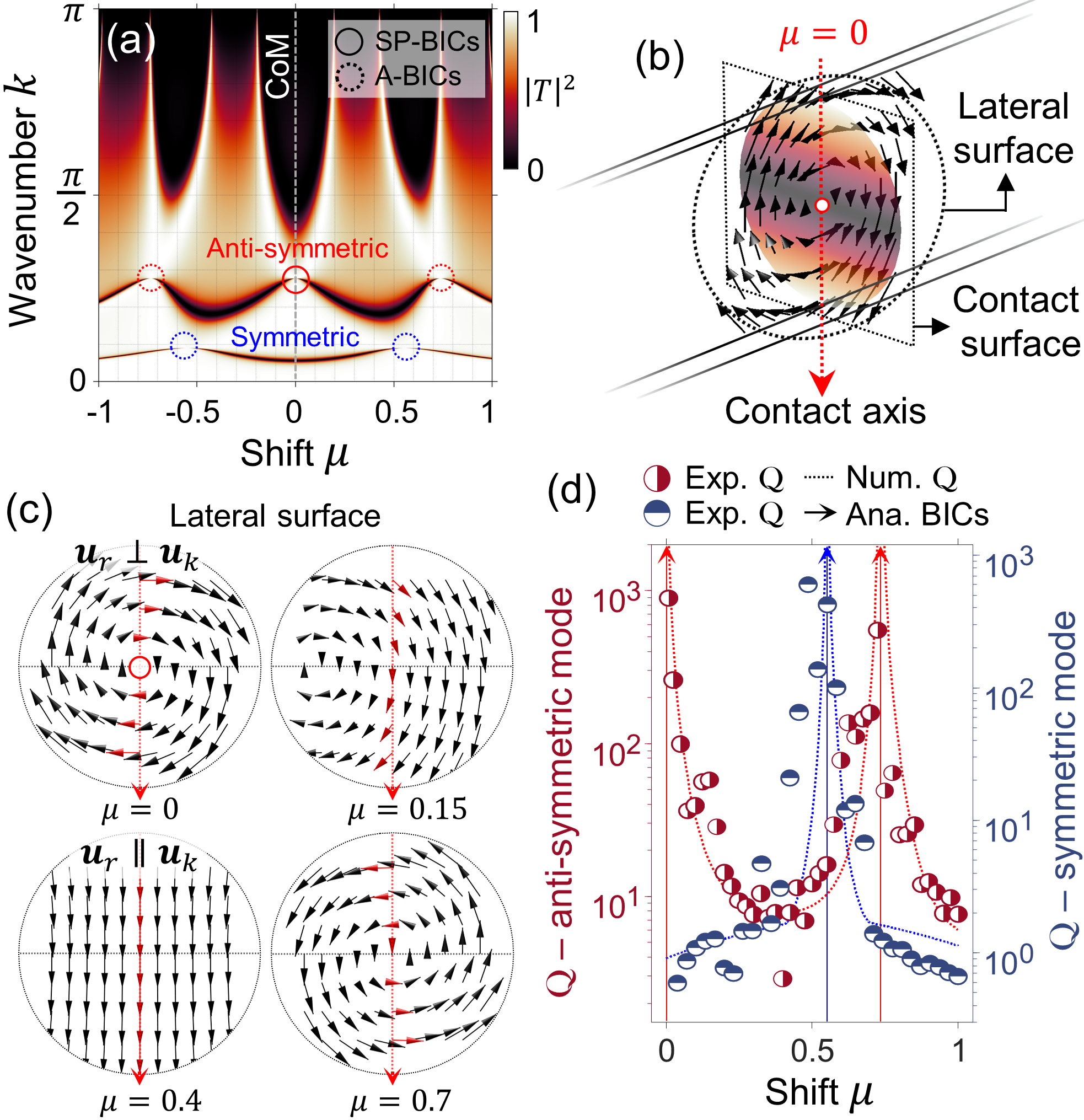}
\caption{polarization-protected BICs. (a) The transmission coefficient for the scattering of a plane wave in a chain with BIC defect as a function of the $k$ and $\mu$. The marked circle (where the coefficient width vanishes) indicates the signatures of BICs. The blue lines represent symmetric modes, while the red lines indicate antisymmetric modes; solid lines denote symmetry-protected BICs (SP-BICs), and dashed lines represent accidental BICs (A-BICs). (b) Displacement field of the BIC defect cylinder at $\mu$=0, one of the marked red circles in (a). The field exhibits a purely rotational mode, orthogonal to both the resonator’s contact boundary and its axis. (c) Displacement field of the BIC defect cylinder on its lateral surface for a given $\mu$. This demonstrates the evolution of polarization alignment between $\mathbf{u}_r$ (displacement vector of the resonator) and $\mathbf{u}_k$ (displacement along the contact axis). (d) Experimental and numerical Q-factors for both symmetric and antisymmetric modes.}
\label{fig2}
\end{figure}

\begin{figure}
\includegraphics [width=0.49\textwidth]{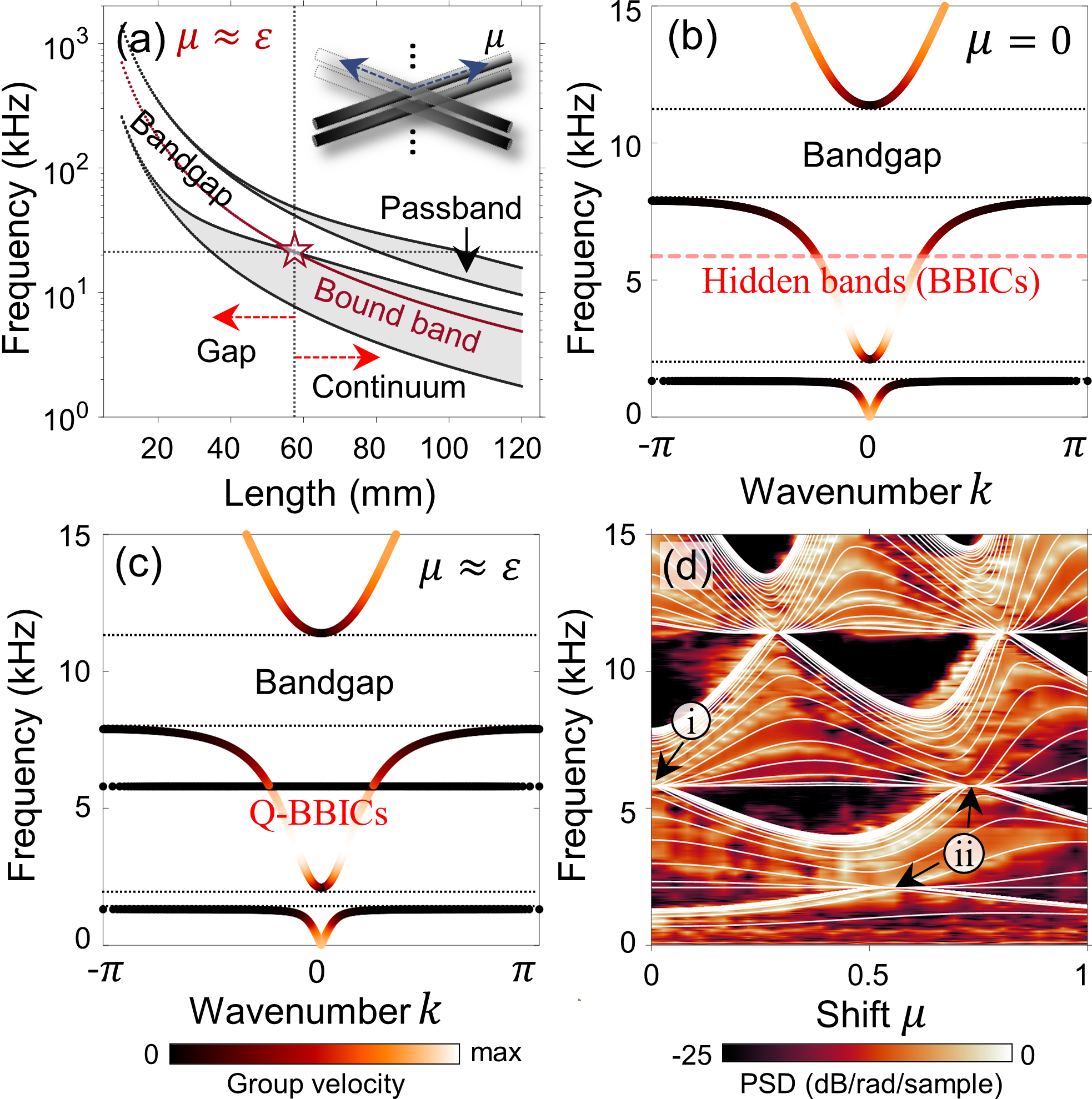}
\caption{Bound bands in the continuum (BBIC) (a) Evolution of bound bands in a periodic chain when $\mu\approx\varepsilon$ for all cylinders, as a function of cylinder length. The bound bands can reside either within the bandgap (white area) or the passband (gray area).
(b)-(c) Band structures for a cylinder chain with a length of 110 mm at $\mu$=0 and $\mu\approx\varepsilon$, respectively. At $\mu$=0, the bound band from antisymmetric modes shows hidden eigenstates (i.e., symmetry-protected), which evolve into a symmetry-broken state at $\mu\approx\varepsilon$, presenting as a quasi-flat band. (d) Experimentally measured frequency response of a 110 mm homogeneous cylinder chain as the overall shift $\mu$ increases. The white lines represent the bulk spectrum calculated using the DEM.}
\label{fig3}
\end{figure}

\begin{figure}
\includegraphics [width=0.49\textwidth]{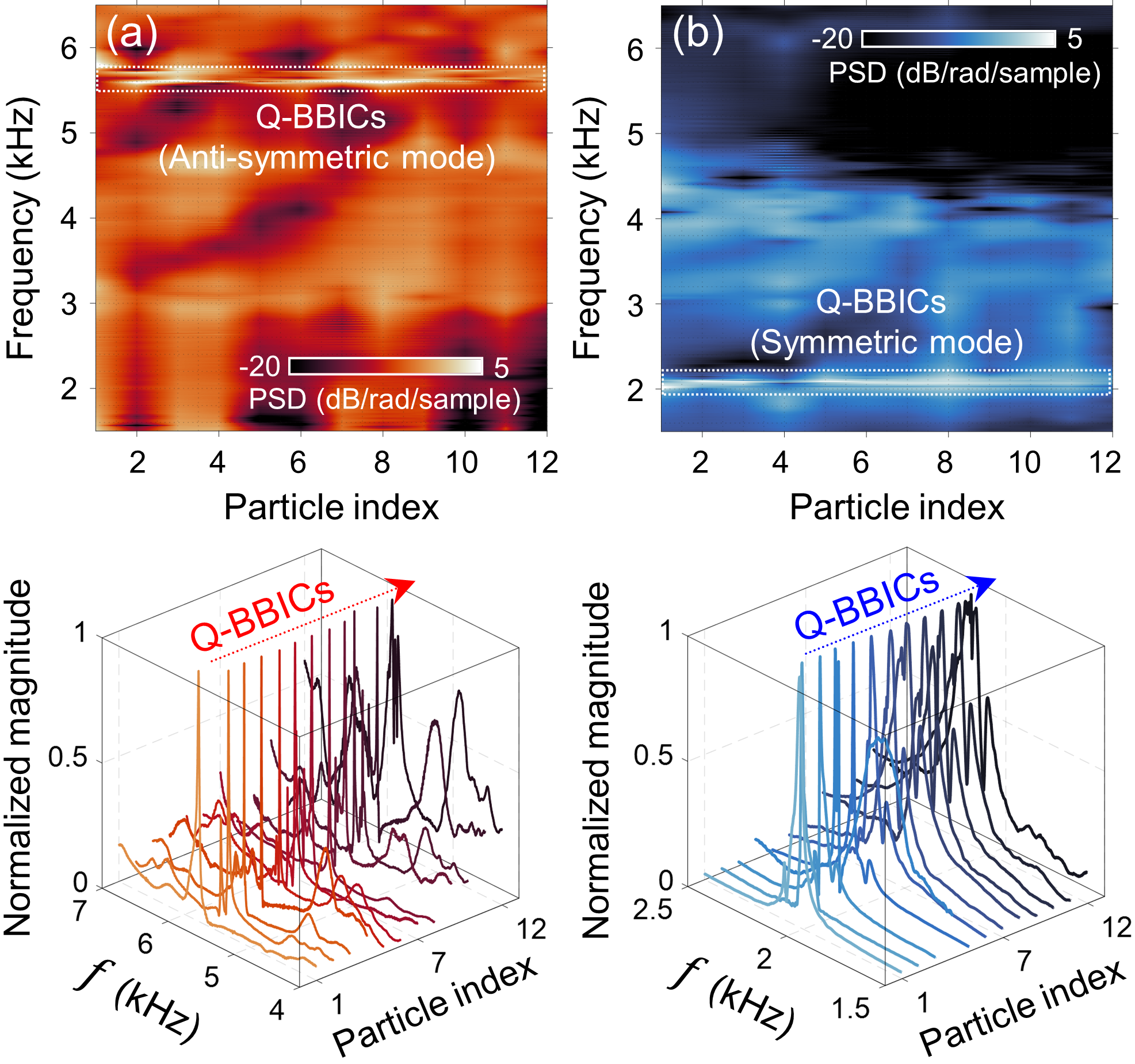}
\caption{Experimental realization of the BBIC formed by a quasi-flat band: (a) for anti-symmetric modes and (b) for symmetric modes (highlighted in white boxes). The colors represent the power spectral density. Each localized flat band exists within a continuous passband. The black regions indicate the bandgap. At the bottom of the spectrum, the flat spectrum evolution marked by a white box is shown.}
\label{fig4}
\end{figure}

\textit{polarization-protected BICs.}---
The incident plane wave results in transmitted ($T$) and reflected ($R$) waves due to its interaction with the defect resonator.
In Fig.~\ref{fig2}(a), we show the transmission coefficients as functions of wavenumber $k$ and shift $\mu$.
The $k$ derived from the relationship $m_0\omega^2=2\beta_c[1-\cos({k})]$ in the acoustic branch of a monatomic chain with host cylinders.
One can see numerous points where the width of the coefficient vanishes, which are the unambiguous signatures of the BICs~\cite{Kim1999,Marinica2008}.
The lower two BIC points arise from the symmetric mode (first bending mode), while the upper three BIC points originate from the anti-symmetric mode (second bending mode). 
We particularly notice the emergence of the BIC in the anti-symmetric mode when the resonator shift coincides with the CoM ($\mu$=0).
In Fig.~\ref{fig2}(b), we show the displacement field of the BIC defect resonator at $\mu$=0 obtained through finite element analysis.
The contact surface (cross-sectional direction) shows the displacement map, while the lateral surface (axis direction) displays the displacement vector field, representing the resonator's bending vibration.
We observe purely rotational modes of the resonator's internal displacement fields, which are perpendicular to the wave propagation at the contact boundary and its axis.
This occurs because the \textit{centroid} separates rotational motion from translational ones. Thus, the purely rotational oscillation does not exert a force on the surrounding host cylinder through the contact boundary, thus being completely decoupled from the radiation continuum
In other words, if a perturbation is introduced inside the resonator, the energy remains perfectly confined and is not radiated externally, indicating the formation of a genuine BIC.
The existence of this nonradiative state in a single resonator is attributed to polarization-protected BICs, which are unique to the solid resonator as they support both longitudinal and transverse modes.
\\
\indent
This mechanism is closely related to the unresponsiveness at the \textit{node}.
The anti-symmetric transverse modes of a cylindrical resonator, which are odd functions, have nodal points at the CoM.
When excited in the out-of-plane direction at the nodal point, these excitations are orthogonal to the anti-symmetric modes. 
This argument is supported by Fig.~\ref{fig2}(a), where the number of BIC points corresponds to the number of nodal points for the bending mode. 
This means that various accidental BICs (A-BICs) appearing at specific $\mu$ can be analytically determined by finding the zeros of the associated mode shapes.
Indeed, the physics-informed DEM confirms that BICs exhibit \textit{near-zero} coefficients for both mass and stiffness in the associated harmonic oscillator (see Supplemental Material~\cite{supp}).
\\
\indent
In Fig.~\ref{fig2}(c), we show the displacement vector field of the resonator at a particular $\mu$.
We observe an undefined displacement field (marked by a red circle) at the origin of symmetry in the resonator at $\mu$=0. 
This occurs due to the perfect geometric symmetry in all directions within the fully three-dimensional geometry, indicating symmetry-protected BICs ($\mathbf{u}_r$$\bot$$\mathbf{u}_k$); $\mathbf{u}_r$ is the displacement vector of a resonator and $\mathbf{u}_k$ displacement along the contact axis (wave vector).
Increasing the shift induces coupling between the resonator's internal displacement and wave propagation, with strong coupling observed around $\mu$=0.4 (i.e., parallel polarization, $\mathbf{u}_r$$\parallel$$\mathbf{u}_k$). 
At $\mu$=0.7, the vector field exhibits characteristics of an accidental quasi-BIC (i.e., quasi-orthogonal polarization).
\\
\indent
In Fig.~\ref{fig2}(d), we show the Q-factor extracted from experiments and numerical simulations using the DEM.
We observe excellent agreement between the experiments and numerics, with high Q-factors ($\approx$~891) at BICs and a dramatic decrease in Q-factor as we deviate from BICs points, following the $Q\propto\mu^{-2}$ relation~\cite{Koshelev2018} (see Supplementary Materials~\cite{supp}).
To our knowledge, these are the highest Q-factors observed in mechanical analogue BIC systems to date.
It should be noted that our defect system differs from previous studies utilizing Friedrich-Wintgen (FW) BICs~\cite{Friedrich1985} in defect layers. 
Our approach leverages polarization-protected BICs that emerge within single-particle resonators, in contrast to FW BICs, which rely on complete destructive interference between modes originating from the defect and the host (i.e., mixing polarizations)~\cite{GomisBresco2017,Timofeev2018,Quotane2018,Pankin2020}.
\vspace{2mm}
\\
\indent
\textit{Bound bands in the continuum (BBICs).}---
Following this finding, a natural question arises: Can the bound “states” through the contact boundary in a single resonator be extended to a bound “band” in periodic systems? Regarding this, we reveal the emergence of bound bands in the continuum as a form of flat bands.
The notable aspect is that this spectrum guarantees both high-Q and dispersionless resonance.
\\
\indent
The coexistence of bound and continuous bands depends on the configuration of the cylindrical resonator. 
We examine how length variations influence this coexistence, as shown in Fig.~\ref{fig3}(a), which shows the evolution of bound bands (red line) from anti-symmetric modes when $\mu\approx\varepsilon$ for all cylinders.
We observe that bound bands either reside within the bandgap (white area) or the passband (gray area), depending on the length marked with an asterisk.
When the bound band lies within the bandgap, it cannot interact with propagating waves and represents a trivial localized mode. 
In contrast, a bound band within the passband can interact with propagating modes, leading to energy accumulation from external perturbations (i.e., resulting in high-Q).
\\
\indent
In Fig.~\ref{fig3}(b)-(c), we present the band structure for a cylinder chain with a length of 110 mm at $\mu$=0 and $\mu\approx\varepsilon$, respectively.
At $\mu$=0, the bound bands from anti-symmetric modes are undefined due to their symmetry-protected configurations (i.e., hidden bands or BBICs). 
With a very small shift ($\mu\approx\varepsilon$), which breaks the resonator's symmetry, these bands evolve into observable Q-BBICs, appearing as flat bands [Fig.~\ref{fig3}(c)].
\\
\indent
Indeed, \textit{flat band} physics has recently gained attention across various fields due to the presence of completely degenerate energy bands~\cite{Leykam2018,Ramachandran2018}. 
Such bands exhibit zero group velocity for all bulk momentum, enabling the existence of compactly localized eigenstates.
Typically, flat bands are achieved by utilizing either internal symmetries or fine-tuned coupling between sites to create complete destructive interference~\cite{Leykam2018}. 
However, it should be noted that in our system, the flat bands originate from BICs from each cylindrical resonator.
In this context, these flat bands ensure a diverging Q-factor.
\\
\indent
The degree of mode coupling can be assessed through the frequency response resulting from an overall shift of the cylinder chain as shown in Fig.~\ref{fig3}(d).
Spectrum plots are obtained experimentally by performing Fourier transformation on the temporal velocity profiles of the chain’s output signal. 
The white lines represent the bulk spectrum calculated from the DEM, which matches well with the experimental results.
As indicated by the arrows, we can clearly observe the near-zero mode coupling and the resulting gap-closing in the spectrum (\circled{i} is symmetry-protected; \circled{ii} is accidental type). At certain shifts, mode coupling becomes significant, leading to avoided crossing phenomena due to the strong coupling between continuous and resonance bands. This results in local resonance bandgaps, which are depicted as black regions.
\\
\indent
To demonstrate the BBICs, we experimentally measure the signal across all cylindrical elements under frequency sweep excitation, synchronizing the spectrum with respect to the particle index. 
In this manner, we detect quasi-flat bands within the continuous bands for anti-symmetric modes [$\mu\approx0$, Fig.~\ref{fig4}(a)] and symmetric modes [$\mu\approx0.55$, Fig.~\ref{fig4}(b)]. 
In the bottom panels, we show the evolution of the flat spectrum for the region marked by the white box in Figs.~\ref{fig4}(a)-(b).
A slight frequency deviation occurs at the first particle due to experimental boundary effects.
As the particles evolve along the chain, minor energy loss is detected due to cumulative material damping.
Despite these influences, the calculated average Q-factor is impressively high, reaching 1028 for the anti-symmetric mode and 428 for the symmetric mode. 
In stark contrast, the average Q-factor for other dispersive structural resonances is just 14 (see the Supplemental Material~\cite{supp} for further details).
\\
\indent
In this way, we validate the proposed BBIC, once again noting that it derives from the capability of a single cylindrical resonator to independently support BICs.
We also note that transitioning from the BBIC in Fig.~\ref{fig4}(a) ($f\approx$ 6 kHz) to the BBIC in Fig.~\ref{fig4}(b) ($f\approx$ 2 kHz) can be easily achieved by simply adjusting the contact boundary. 
This \textit{in situ} tuning feature makes the system flexible without needing to alter other structural parameters.
The high-Q energy confinement throughout the structure indicates its potential as an efficient platform for resonant sensors and energy harvesting devices.
To back up this, we experimentally demonstrate the strong energy accumulation in each cylindrical element under monochromatic excitation at the bound mode’s frequency in the Supplementary Materials~\cite{supp}.
\vspace{2mm}
\\
\indent
\textit{Conclusions}---
In the present work, we have proposed a highly tunable mechanical system made of cylindrical granular particles, which has demonstrated polarization-protected BICs within a contact boundary. 
With \textit{in situ} tuning of the cylindrical resonator's contact boundary, orthogonality between the resonator's internal displacement fields and wave propagation along the contact axis can be realized, resulting in complete localization within the resonator.
These bound modes, arising from single-particle resonators, have extended the bound bands in the continuum within periodic structures composed of such resonators. 
Experimentally, we have observed the emergence of quasi-flat spectra in a finite chain with broken symmetry in each resonator. 
All cylindrical particles have exhibited high-Q and dispersionless resonances.
\\
\indent
This study suggests multiple promising directions for future work. In the context of linear dynamics, the graded BIC concept—derived from the graded profiles of cylinder lengths within a chain—holds particular interest. 
Beyond the linear regime, a key advantage of this system is its ability to trigger nonlinear wave dynamics through contact nature under increased amplitude of dynamic excitation. 
We thus believe that this study offers a promising roadmap for developing an experimentally and theoretically accessible testbed to investigate the interplay between nonlinearity and BICs.
\section*{Acknowledgements}
This work was financially supported by the POSCO-POSTECH-RIST Convergence Research Center program funded by POSCO and the National Research Foundation (NRF) grants (NRF-2019R1A2C3003129, CAMM-2019M3A6B3030637, NRF-2019R1A5A8080290) funded by the Ministry of Science and ICT (MSIT) of the Korean government, and the grant (PES4400) from the endowment project of “Development of smart sensor technology for underwater environment monitoring” funded by Korea Research Institute of Ships Ocean engineering (KRISO). E.K. also acknowledges the support of the National Research Foundation grant (NRF-2020R1A2C2013414) and the Commercialization Promotion Agency for R\&D Outcomes(COMPA) grant funded by the Korean Government (Ministry of Science and ICT, 2023).

\bibliographystyle{apsrev4-2}
\bibliography{Reference.bib}


\end{document}


\title{SUPPLEMENTAL MATERIALS:\\Bound States-to-Bands in the Continuum in Cylindrical Granular Crystals}

 \author{Yeongtae Jang}
     \affiliation{%
     Department of Mechanical Engineering, Pohang University of Science and Technology (POSTECH), Pohang 37673, Republic of Korea}
     \email{jrsho@postech.ac.kr}
 \author{Seokwoo Kim}
     \affiliation{%
     Department of Mechanical Engineering, Pohang University of Science and Technology (POSTECH), Pohang 37673, Republic of Korea}
 \author{Dongwoo Lee}
     \affiliation{%
     Department of Mechanical Engineering, Pohang University of Science and Technology (POSTECH), Pohang 37673, Republic of Korea}
  \author{Eunho Kim}\email{eunhokim@jbnu.ac.kr}
      \affiliation{%
        Division of Mechanical System Engineering, Jeonbuk National University, Jeonju, 54896, Republic of Korea}
      \email{eunhokim@jbnu.ac.kr}
      \affiliation{%
     Department of JBNU-KIST Industry-Academia Convergence Research, Jeonju, 54896, Republic of Korea}
 \author{Junsuk Rho}\email{jrsho@postech.ac.kr}
     \affiliation{%
     Department of Mechanical Engineering, Pohang University of Science and Technology (POSTECH), Pohang 37673, Republic of Korea}
     \affiliation{%
     Department of Chemical Engineering, Pohang University of Science and Technology (POSTECH), Pohang 37673, Republic of Korea}
    \affiliation{%
    Department of Electrical Engineering, Pohang University of Science and Technology (POSTECH), Pohang 37673, Republic of Korea}
     \affiliation{%
     POSCO-POSTECH-RIST Convergence Research Center for Flat Optics and Metaphotonics, Pohang 37673, Korea}
     
\maketitle
\tableofcontents
\newpage
\section{I.~Experimental setup\label{sec:exp}}
\begin{figure}[htp]
\includegraphics [width=0.98\textwidth]{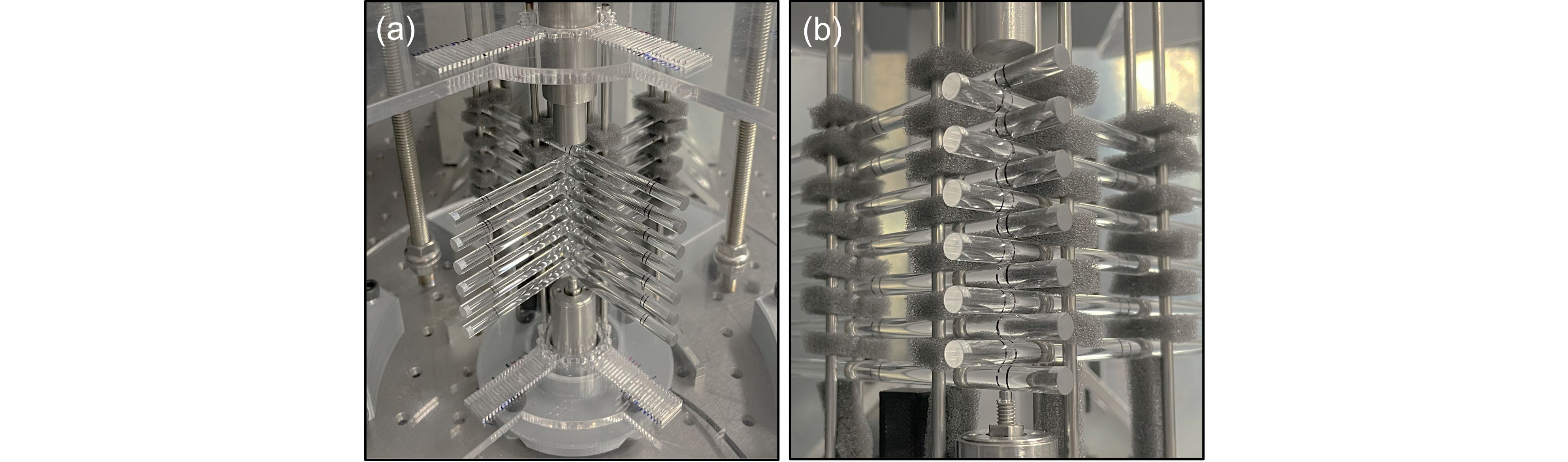}
    \caption{Experimental setup of cylindrical beam crystals. (a) Orthogonally stacked cylinder chain via centroidal contact. (b) A magnified view of a centroidal contact-shifted cylindrical chain.}
\label{figS1}
\end{figure}

Figure~\ref{figS1} displays digital images of the experimental setup constructed in this study.
We investigate two configurations of cylinder chains: a defect-embedded chain for the bound “state” in the continuum (BICs) and a centroid-shifted periodic chain for the bound “band” in the continuum (BBICs). 
Here, we specifically show a centroid-shifted homogeneous chain comprising long cylindrical elements, while the defect-embedded chain involves placing a long cylinder in the middle of a chain composed of shorter cylinders.
\\
\indent
As noted in the main text, an \textit{in situ} shift enables us to tune the strength of local resonance coupling, stemming from the bending modes of the cylindrical beam [see Fig.~\ref{figS1}(b)].
The beams are made from fused quartz (a density~$\rho=2187~\mathrm{kg/m^3}$, a Young's modulus~$E=72~\mathrm{GPa}$, and a Possion's ratio~$\nu=0.17$), which is characterized by low material damping.
Thus, we can capture the dynamics in the cylinder chain more clearly with minimal wave dissipation.
The chains are supported by vertically positioned guide rods,  which aid the out-of-plane motion of the cylinder and prevent structural buckling. To align the rods parallel along the chain, we place soft polyurethane foam between the cylinders. 
Due to the foam's stiffness being significantly lower than the rigidity of the cylinder and the contact stiffness between the cylindrical rods, its impact on the cylinder's motion is negligible.
The experimental process and data acquisition are as follows: A piezo-actuator (Piezomechanik PSt 500\textbackslash10\textbackslash25) excites the bottom of the cylinder chains with a frequency sweep signal or harmonic excitations at a BIC frequency.
A 13N free weight positioned atop the chain provides pre-compression ($F_0$), thereby maintaining the system dynamics within the linear regime.
The localized velocity profiles near the contact site are detected using non-contact laser Doppler vibrometry mounted on an automatic guide rail.
\newpage
\section{II.~Discrete Element Model\label{sec:dem}}
\begin{figure}[htp]
\includegraphics [width=0.90\textwidth]{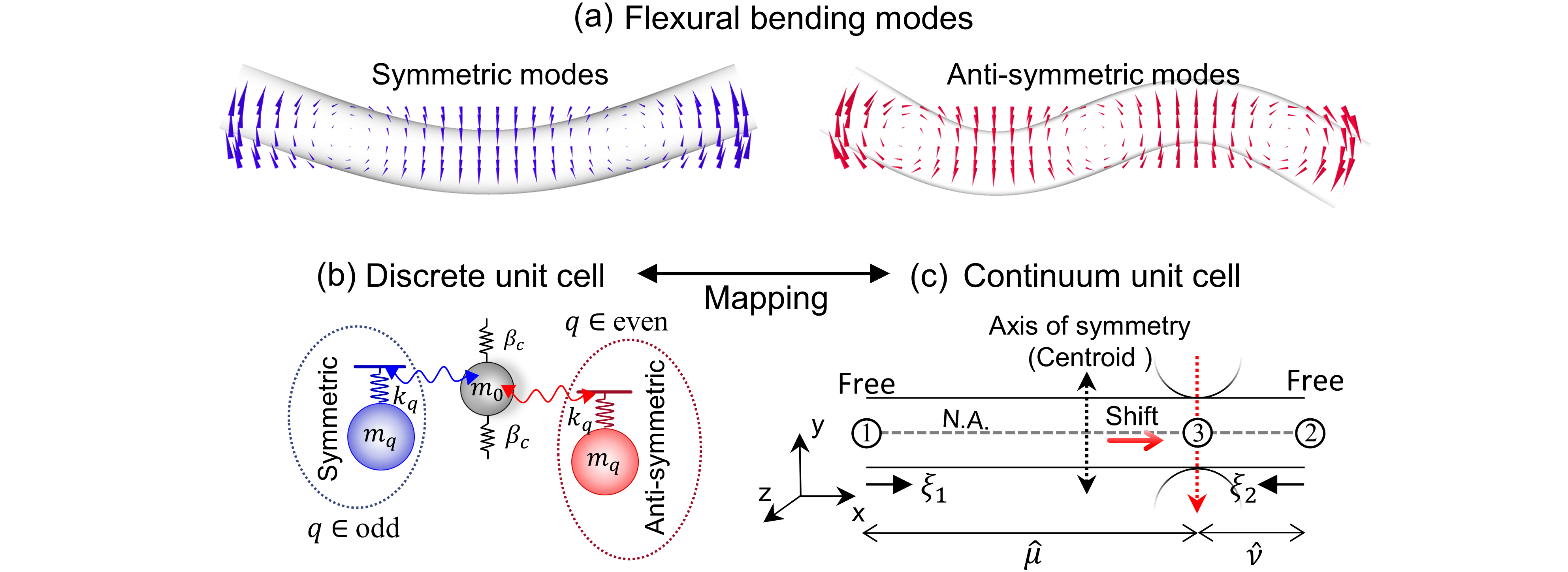}
    \caption{(a) Bending mode shapes of the cylindrical resonators for the symmetric and anti-symmetric modes. The arrow cones denote the displacement fields of each bending mode. (b) Discretized unit cell, where mode coupling from the bending modes is represented by internal resonators with mechanical coefficient $m_q$ and $k_q$. (c) Schematic of the continuum beam and its boundary conditions. This continuum unit cell is directly mapped to the discretized unit cell using physics-informed discrete element modeling~\cite{Jang_2024}.}
\label{figS2}
\end{figure}

In this section, we develop a discrete element model (DEM) for the continuous cylindrical elements using the wave physics-informed modeling approach recently introduced in Ref.~\cite{Jang_2024}.
These reduced-order models are widely utilized in condensed matter physics to simplify wave dynamics computations across periodic and aperiodic structures, providing intuitive insights into underlying wave physics~\cite{Matlack2018}.
Especially, the proposed DEM modeling approach includes local resonance coupling, stemming from the interaction between flexural bending modes of the continuum elements [refer to Fig.~\ref{figS2}(a)] and propagating waves within the chains.
Flexural bending modes of the cylindrical element are classified into symmetrical (odd-numbered) and anti-symmetrical (even-numbered) modes, where we particularly depict the first and second modes.
The arrow cones denote the displacement fields of each bending mode.
In our contact-based system, wave propagation occurs through contacts between neighboring cylinders, thus the strength of wave coupling with each vibrational bending mode depends on the displacement field magnitude at the contact location. 
Therefore, shifting the contact location allows for straightforward tuning of local resonance coupling strength.
\vspace{2mm}
\\
\indent
The contact force ($F$) between neighboring cylinders is governed by the nonlinear Hertz law~\cite{johnson1987contact}, represented as $F=A\delta^{3/2}$, where $\delta$ is the compressive displacement.
The stiffness coefficient ($A$) depends on the geometries, material properties, and contact angle ($\theta$) between cylindrical elements.
For a contact angle of $\theta=90^{\circ}$, the stiffness coefficient simplifies to $A=2E\sqrt{R}/3(1-v^2)$; where, $R$ is the radius of the cylinders, $E$ and $v$ are Young's modulus and Poisson ratio.
By assuming that external dynamic forces are significantly smaller than the static forces ($F_0$) within the cylinder chains, we can linearize the contact stiffness ($\beta_c$) as follows:
\begin{equation}
    \beta_c=\frac{3}{2}A^{2/3}F_0^{1/3}.
\end{equation}
\\
\indent
For constructing the DEM, it is essential to determine the parameters of a unit cell through its discretized spring and mass coefficients [e.g., $m_0$, $m_q$, and $k_q$ as shown in Fig.~\ref{figS2}(b), where $q$ denotes the mode number]. 
These parameters are derived by correlating the analytic continuum dynamics [Fig.~\ref{figS2}(c)] with the discretized model. 
This mathematical modeling process is systematically divided into three main stages:
\vspace{2mm}
\\
\indent
1.~We begin with the dispersion relationship $\omega(k)$ of a discretized unit cell. 
In this dispersion curve, we can formalize the specific solution where the group velocity (${\partial{\omega}}/{\partial{k}}$) is zero despite the unknown spring-mass coefficients of a unit cell. 
Since this solution is in the form of a standing wave, we can focus on the \textit{vibration} state within a single unit cell, not on traveling waves.
Through a series of algebraic equations of this standing wave solution, we can establish the inner product of a matrix.
\vspace{2mm}
\\
\indent
2.~By focusing on the vibration state, we can straightforwardly apply continuum dynamics to the single cylinders. 
The dynamics of the continuum beam are analyzed using well-established beam theory, governed by a partial differential equation (PDE), which requires boundary conditions to determine dynamic quantities such as eigenfrequencies and eigenmodes. 
In our contact-based systems, these boundary conditions rely on “contact behavior,” resulting from the wavenumber-dependent movement of cylinders. Therefore, by applying the physical contact behavior of the wavenumber with zero group velocity as a boundary condition, we determine multiple eigenfrequencies.
\vspace{2mm}
\\
\indent
3.~Finally, by applying the multiple eigenfrequencies obtained in stage 2 to the elements of the matrix obtained in stage 1, we can analytically determine the discretized parameters of the spring-mass system. This approach incorporates various factors such as the cylinder's geometry and material properties, as well as the effect of the shift in contact location.
Connecting the discretized unit cell via contact stiffness allows us to assemble the complete DEM. This model system can be readily extended to various wave problems (e.g., scattering problem, time evolution, etc) due to its simplified ordinary differential equations (ODE) formulation.
\vspace{2mm}
\\
\indent
As mentioned earlier, we first begin with the dispersion relation of the discrete unit cell.
Within the homogeneous chains, the equations of motion for the $n$-th unit cell is
\begin{subequations}
\begin{eqnarray}
    &m&_{n,0}\ddot{u}_{n,0} = {\beta}_c\left( u_{n-1,0} - 2 u_{n,0} + u_{n+1,0} \right) + \sum_{q} k_{n,q}(u_{n,q} - u_{n,0}),\label{eq:S2a}  \\    
    &m&_{n,q} \ddot{u}_{n,q} = k_{n,q}( u_{n,0} - u_{n,q}  ), \quad q \in \mathbb{Z}^{+}.\label{eq:S2b}
\end{eqnarray}
\end{subequations}
Here $u_{n,0}$ and $u_{n,q}$ are the displacement of a primary mass $m_0$ and the local mass $m_q$ (with the $q$-th mode), respectively.
Assuming harmonic motion and taking $u_{n,0}=\bm{U}_0\text{exp}(i\omega{t})$ and $u_{n,q}=\bm{U}_q\text{exp}(i\omega{t})$ in Eq.~(\ref{eq:S2b}), we derive the following relation:
\begin{equation}
\begin{aligned}
    u_{n,q} = \frac{\omega_q^2}{\omega_q^2 - \omega^2}u_{n,0}, \quad q \in \mathbb{Z}^{+},\label{eq:S3}
\end{aligned}
\end{equation}
where $\omega_{q}=\sqrt{k_q/m_q}$ is the resonant frequency of an internal resonator and $q$ is the mode number associated with flexural bending modes, $q=1, 3, 5...$ for the symmetric modes, and $q=2, 4, 6...$ for the anti-symmetric modes.
By substituting this relation into Eq.~(\ref{eq:S2a}), one can obtain
\begin{equation}
\begin{aligned}
    -m_{\text{eff}} \omega^2 u_n = {\beta}_c \left( u_{n-1} - 2u_n + u_{n + 1}\right), \label{eq:S4}
\end{aligned}
\end{equation}
where
\begin{equation}
\begin{aligned}
    m_{\text{eff}}(\omega) = m_0 + \sum_{q} m_q\frac{\omega_q^2}{\omega_q^2 - \omega^2}.
\end{aligned}
\end{equation}
The universal index 0 for displacement is omitted here.
Explicitly, Eq.~(\ref{eq:S4}) represents the motion of conventional monatomic chains, with an effective mass that depends on the excitation frequency $\omega$.
We can then derive the dispersion relation by applying Floquet's theorem, expressed in the form $u_{n\pm{1}}=u_n\text{exp}(\mp{ika})$ to Eq.~(\ref{eq:S4}).
\begin{equation}
    ka=\cos^{-1}\left(1-\frac{m_\text{eff}\omega^2}{2\beta_c}\right),
    \label{eq:S6}
\end{equation}
where $ka$ is the normalized wavenumber with the lattice constant $a$.
\begin{figure}
\includegraphics [width=0.98\textwidth]{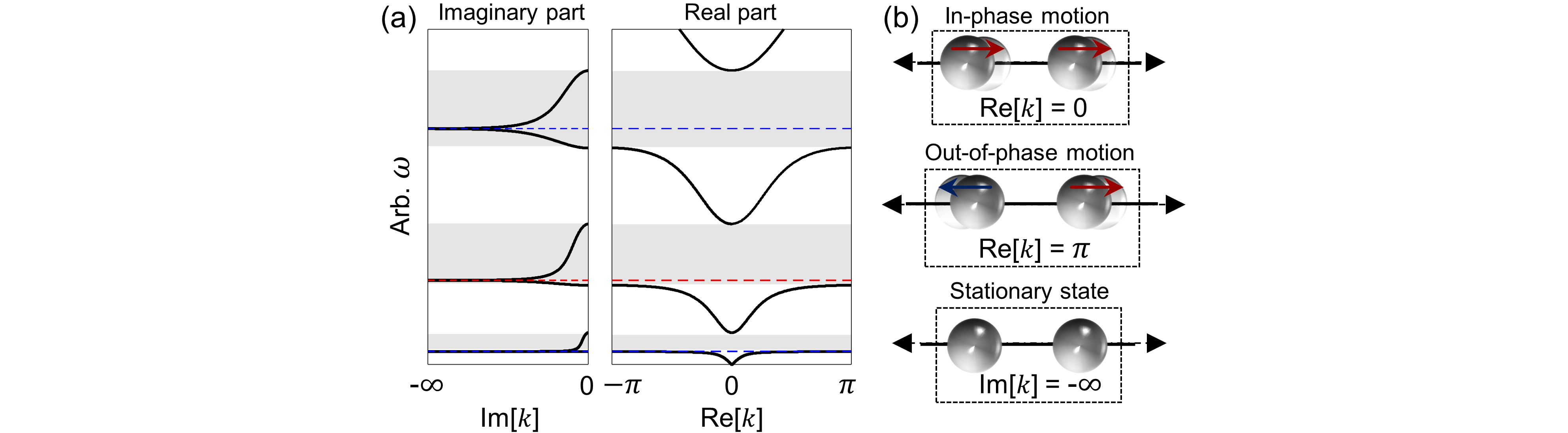}
    \caption{(a) Band structure of multiple locally resonant chains as a function of the complex wavenumber. The left side shows the imaginary part, while the right side displays the real part, with the gray area indicating the local resonance bandgap. (b) Relative motion between unit cells at specific wavenumbers, representing standing or semi-standing waves.}
\label{figS3}
\end{figure}
\\
\indent
In Fig.~\ref{figS3}(a), we show the generic dispersion curve for our multiple locally resonant chains with the normalized wavenumber ($k$) decomposed into a complex value.
The real part is displayed on the right, and the imaginary part on the left.
The gray region indicates the local resonance-induced bandgap, where the wave solution appears in the imaginary part of the dispersion curve, exhibiting singularities. 
The blue dotted lines represent symmetrical modes, and the red dotted lines represent anti-symmetrical modes.
\\
\indent
It should be noted here that our 1D cylinder chain does not include coupling effects between different local resonance modes such as veering and locking phenomena~\cite{Mace2012} due to mono-channel wave transport. 
As a result, only the coupling between the propagating wave and multiple local resonances―referred to as avoided crossing―is considered. 
This explanation also justifies our choice of the mass-with-mass configuration.
\\
\indent
As noted in the previous modeling process (Stage 1), we seek the frequencies with zero group velocity in the dispersion curve for mapping to the continuum dynamics of a single cylindrical beam. 
We categorize two types of standing waves in the real part of the dispersion curve and one type of semi-standing waves in the imaginary part.
\vspace{2mm}
\\
\indent
(i)~\textit{Onset frequency} ($\omega^{s}_{p}$ at $\text{Re}[k]=0$):~At a wavenumber of zero, neighboring cells exhibit in-phase motion because the wavelength becomes infinite. This condition eliminates force interactions between adjacent cells, leading each cell to oscillate at its own resonant frequencies. Therefore, all these resonant states (except for static regimes at $\omega=0$) have zero group velocity.
\vspace{2mm}
\\
\indent
(ii)~\textit{Cutoff frequency} ($\omega^{c}_{p}$ at $\text{Re}[k]=\pi$):~At the Brillouin zone boundary point, neighboring cells exhibit out-of-phase motion, leading to force interactions between them.
Mathematically, considering plane waves, $u_n=\bm{A}e^{i(kn-\omega{t})}$ within a monoatomic chain, $u_n$ is reduced $\bm{A}e^{i\pi{n}}e^{\i\omega{t}}$ when $k=\pi$.
Thus, $e^{i\pi{n}}$ simplifies to $(-1)^n$, indicating that there is only a phase difference between neighboring particles. 
As a result, there is no energy flow within the chain, which corresponds to zero group velocity.
\vspace{2mm}
\\
\indent
(iii)~\textit{Local resonance frequency} ($\omega_{q}$ at $\text{Im}[k]=-\infty$):~In the imaginary part of the dispersion curve, we focus on singularities where energy transport is completely prohibited.
This prohibition arises from the exact anti-phase motion of the local resonator against the external wave. 
In terms of mathematics, if we introduce a complex form for the wavenumber, expressed as $k=\alpha+i\beta$, the displacement of the particles becomes proportional to $\bm{A}e^{-\beta(n/a)}$.
Therefore, when the coefficient $-\beta$ becomes infinite, the wave amplitude approaches zero. 
In this context, energy flow within the chain is restricted, but each particle oscillates with its own local resonance frequency $\omega_q$. 
From the perspective of an effective particle, the mass is in a stationary (pinned) state.
Therefore, this state can also be regarded as a semi-standing wave.
\vspace{2mm}
\\
\indent
In Fig.~\ref{figS3}(b), we show the schematic of the relative motion between particles in the various aforementioned standing wave states.
\\
\indent
By substituting $k=0$ with $\omega\rightarrow\omega^{s}_{p}$ and $k=\pi$ with $\omega\rightarrow\omega^{c}_{p}$ into the dispersion relation (Eq.~\ref{eq:S6}), we can obtain
\begin{subequations}
\begin{eqnarray}
    \left(m_0 + \sum_{q}{m_q} \frac{\lambda_q}{\lambda_q - \lambda_p^s}\right)\lambda_p^s&=&0,\quad\forall p \geq 0,\label{eq:S7a}  \\    
    \left(m_0 + \sum_{q}{m_q} \frac{\lambda_q}{\lambda_q - \lambda_p^c}\right)\lambda_p^c&=&4\beta_c,\quad\forall p \geq 0,\label{eq:S7b}
\end{eqnarray}
\end{subequations}
To simplify calculations, we express $\omega^2$ in terms of the eigenvalues $\lambda$.
In the first branch, the onset frequency is zero because our system inherently includes rigid body motion.
We treat this static regime independently, emphasizing that its associated frequency index is zero ($p=0$).
Thus, the zero frequency makes Eq.~(\ref{eq:S7a}) self-evident, and from Eq.~(\ref{eq:S7b}), we obtain
\begin{equation}\label{eq:S8}
\begin{gathered}
    m_0 + \sum_{q}{m_q} \frac{\lambda_q}{\lambda_q - \lambda_0^c} = 4\frac{{\beta}_c}{\lambda_0^c}.
\end{gathered}
\end{equation}
For the higher branch ($p\geq1$), we can construct a unified equation by subtracting Eq.~(\ref{eq:S7a}) from Eq.~(\ref{eq:S7b}) for the same mode number.
\begin{equation}
    m_0 + \sum_{q}{m_q} \frac{\lambda_q(\lambda^c_p-\lambda^s_p)}{(\lambda_q - \lambda_p^c)(\lambda_q - \lambda_p^s)} = 4\frac{{\beta}_c}{\lambda_p^c},\quad\forall p \geq 1.
    \label{eq:S9}
\end{equation}
To simplify the expression, we introduce the inverse-reduced eigenvalue $\Omega$, which represents the difference between the inverse eigenvalues associated with the onset, cutoff, and local resonance frequencies.
\begin{equation}\label{eq:S10}
\begin{gathered}
    \Omega_{p,q}^{s(c)} \equiv  \frac{1}{\lambda_p^{s(c)}} - \frac{1}{\lambda_q}. 
\end{gathered}
\end{equation}
With this inverse-reduced eigenvalue, we form a matrix inner product by merging Eq.~(\ref{eq:S8}) and Eq.~(\ref{eq:S9})
\begin{equation}\label{eq:S11}
\begin{gathered}
\langle\bm{\widehat{\Lambda}}_{p,q},\bm{\mathrm{M}}_p\rangle =\bm{\mathrm{K}}_q, \quad  \text{for } p, q \in \{1, 2, \ldots, M\},
\end{gathered}
\end{equation}
where
\begin{equation}
    \begin{aligned}
    \mathbf{\hat{\Lambda}}_{p,q} = \frac{1}{\lambda_q}\left[\frac{1}{\Omega_{p,q}^c} - \frac{1}{\Omega_{p,q}^s}   \right]. \label{eq:S12}
    \end{aligned}
\end{equation}
Here, $M$ represents the total number of modes considered in the system; the $\mathbf{\hat{\Lambda}}$ is the dimensionless eigenmatrix, and $\mathbf{M}$ and $\mathbf{K}$ are column vectors associated with the mass distribution of unit cells and contact stiffness between unit cells, respectively.
\vspace{3mm}
\\
\indent
We now proceed to stage 2, where we investigate the frequencies of standing waves ($\omega^s_p,~\omega^c_p,~\omega_q$) using continuum mechanics (beam theory).
Among various beam theories, we adopt the Euler-Bernoulli beam theory, which is particularly suitable for analyzing beams with high aspect ratios. 
In the absence of external forces, the governing equation of the Euler-Bernoulli beam is expressed as follows:
\begin{equation}\label{eq:S13}
    \rho\mathrm{A}\frac{\partial^2 w(x,t)}{\partial t^{2}} + \frac{\partial^2}{\partial x^{2}}\left[EI\frac{\partial^2 w(x,t)}{\partial x^{2}}\right] = 0,
\end{equation}
where $\rho$ is the density, $A$ is the cross-sectional area, $E$ is Young's modulus, and $I$ is the area moment of inertia about the $z$-axis.
The deflection $w(x,t)$ in the $y$-direction arises from the transverse vibration of the beam.
For a harmonic solution, let $w(x,t)=W(x)\text{exp}(i\omega{t})$, leading to the following expression from Eq.~(\ref{eq:S13}):
\begin{equation}\label{eq:S14}
    W_{\xi \xi \xi \xi}(\xi)-\psi^4(\xi) = 0,
\end{equation}
where the subscript $\xi$ denotes differentiation with respect to the argument $W$. 
For computational convenience, we utilize a non-dimensional variables, $\xi=x/L$, and a variable $\psi^4=({\rho{A}\omega^2L^4})/({EI})$; here, $L$ represents the length of the beam.
As shown in Fig.~\ref{figS2}(b), our beam exhibits a free-free boundary condition at the edges and an internal contact boundary condition that depends on the shift $\hat{\mu}$.
Therefore, based on the contact location, we can consider the beam equation as two separate parts, $\hat{\mu}~(0\leq\hat{\mu}\leq{1})$ and $\hat{\nu}~(\hat{\nu}=1-\hat{\mu})$, i.e., a double-span beam.
Assuming a solution to Eq.~(\ref{eq:S14}) of the form $A\text{exp}(\sigma\xi)$, the general solution of the equation can be calculated as:
\begin{equation}\label{eq:S15}
    W_{i}(\xi_{i}) = a_{i}\cosh{\psi\xi_{i}} + b_{i}\sinh{\psi\xi_{i}} + c_{i}\cos{\psi\xi_{i}} + d_{i}\sin{\psi\xi_{i}}, \quad \forall~i=1,2,
\end{equation}
where, $\xi_{i}$ denote the local coordinate for each beam.
We note that since our stacked cylinder chain exhibits mirror symmetry about the cylinder's centroid (center-of-mass position). 
Therefore, in the main text, we simplify the analysis by setting the shift $\mu$ from 0 to 1, covering the range from the beam's center to its end.
In this beam theory, we consider a coordinate system for the entire beam length, resulting in the relationship $\mu=2(\hat{\mu}-0.5)$.
\\
\indent
To determine the unknown constants ($a_i,b_i,c_i,d_i,~i\in[1,2]$), we require eight boundary conditions. 
Initially, we can apply four boundary conditions at the beam's edges ($\xi_1=\xi_2=0$), expressed as:
\begin{equation}
\label{eq:S16}
W_{1(2),\xi\xi}(0)=0,\quad W_{1(2),\xi\xi\xi}(0)=0.
\end{equation}
The remaining boundary conditions result from the four patching conditions related to the contact behavior at the contact point ($\xi_1=\hat{\mu},~\xi_2=\hat{\nu}$).
For the various standing waves discussed earlier, this contact behavior depends on the assigned wavenumber (i.e., wavenumber-dependent boundary conditions). 
Revisiting the standing waves, the physical boundary conditions arising from the relative particle motion [see Fig.~\ref{figS3}(b)] at the contact point are as follows:
\vspace{2mm}
\\
\indent
(i)~\textit{Onset frequency} ($\omega^{s}_{p}$ at $\text{Re}[k]=0$):~Due to their in-phase motion from the long-wavelength scales, the interaction between neighboring cylinders is prevented. 
Thus, we impose the continuity of the structural response (e.g., displacement, slope, bending moment, and shear force) at the contact point (i.e., compatibility condition).
\begin{equation}\label{eq:S17}
    \begin{split}
        W_1(\hat{\mu})-W_2(\hat{\nu})=0,~W_{1,\xi}(\hat{\mu})+W_{2,\xi}(\hat{\nu})=0,~W_{1,\xi\xi}(\hat{\mu})-W_{2,\xi\xi}(\hat{\nu})=0,~W_{1,\xi\xi\xi}(\hat{\mu})+W_{2,\xi\xi\xi}(\hat{\nu})=0.
     \end{split}
\end{equation}
By substituting the general solution (Eq.~\ref{eq:S15}) into the edge (Eq.~\ref{eq:S16}) and contact (Eq.~\ref{eq:S17}) boundary conditions, we derive the following characteristic equation ($\mathcal{C}$) for the onset frequency:
\begin{equation}\label{eq:S18}
    \mathcal{C}_s(\psi):~\cos{\psi}\cosh{\psi}-1=0.
\end{equation}
This characteristic equation can be solved numerically, for instance, using Newton's method, yielding multiple onset frequencies $\omega_p^s$.
\vspace{2mm}
\\
\indent
(ii)~\textit{Cutoff frequency} ($\omega^{c}_{p}$ at $\text{Re}[k]=\pi$):~The out-of-phase motion between adjacent cylinders leads to maximum interactions in terms of contact force. 
In continuum mechanics, this contact force is equivalent to the shear force resulting from the pure bending of a beam. 
Thus, we can model the contact locations as elastic supports. 
Along with ensuring continuity of other structural responses (i.e., displacement, slope, and bending moment), we apply the following contact boundary condition:
\begin{equation}\label{eq:S19}
    \begin{split}
        W_1(\hat{\mu})-W_2(\hat{\nu})=0,~W_{1,\xi}(\hat{\mu})+W_{2,\xi}(\hat{\nu})=0,~W_{1,\xi\xi}(\hat{\mu})-W_{2,\xi\xi}(\hat{\nu})=0,~W_{1,\xi\xi\xi}(\hat{\mu})+W_{2,\xi\xi\xi}(\hat{\nu})=-2\hat{\beta}_{c}W_1(\hat{\mu}),
     \end{split}
\end{equation}
where $\hat{\beta}_c$ is a dimensionless contact stiffness defined as $\frac{\beta_c{L^3}}{EI}$ between the neighboring cylindrical elements.
With free boundary conditions at the beam's edges, we obtain the following characteristic equation:
\begin{equation}\label{eq:S20}
    \begin{split}
    \mathcal{C}_c(\psi):~\psi^3(\cos{\psi}\cosh{\psi}-1)-2\hat{\beta_c}((1 +& \cos{\psi \hat{\mu}} \cosh{\psi \hat{\mu}}) (\cos{\psi \hat{\nu}} \sinh{\psi \hat{\nu}}    - \sin{\psi \hat{\nu}} \cosh{\psi \hat{\nu}})\\ + 
    &(1 + \cos{\psi \hat{\nu}} \cosh{\psi \hat{\nu}}) (\cos{\psi \hat{\mu}} \sinh{\psi \hat{\mu}} - \sin{\psi \hat{\mu}} \cosh{\psi \hat{\mu}}))=0.
    \end{split}
\end{equation}
The nontrivial solution to this characteristic equation yields the multiple cutoff frequency $\omega^c_p$.
\vspace{2mm}
\\
\indent
(iii)~\textit{Local resonance frequency} ($\omega_{q}$ at $\text{Im}[k]=-\infty$):~
Stationary states are well-known for \textit{vibration isolation} in classical vibration theory~\cite{inman1994}. Therefore, we can apply intermediate pinned or anti-symmetric boundary conditions at the contact location.
\begin{equation}\label{eq:S21}
    \begin{split}
        W_1(\hat{\mu})=0,~W_2(\hat{\nu})=0,~
        W_{1,\xi}(\hat{\mu})+W_{2,\xi}(\hat{\nu})=0,~~W_{1,\xi\xi}(\hat{\mu})-W_{2,\xi\xi}(\hat{\nu})=0.
    \end{split}
\end{equation}
Similarly, we derive the characteristic equation for the local resonance frequency.
\begin{equation}\label{eq:S22}
    \begin{split}
    \mathcal{C}_\ell(\psi):~(1 + \cos{\psi \hat{\mu}} \cosh{\psi \hat{\mu}}) &(\cos{\psi \hat{\nu}} \sinh{\psi \hat{\nu}}    - \sin{\psi \hat{\nu}} \cosh{\psi \hat{\nu}})\\ + 
    &(1 + \cos{\psi \hat{\nu}} \cosh{\psi \hat{\nu}}) (\cos{\psi \hat{\mu}} \sinh{\psi \hat{\mu}} - \sin{\psi \hat{\mu}} \cosh{\psi \hat{\mu}})=0.
    \end{split}
\end{equation}
Remarkably, we observe that the three different expressions for the characteristic equation satisfy the following relation:
\begin{equation}\label{eq:S23}
    \mathcal{C}_c(\psi)=\psi^3\mathcal{C}_s(\psi)-2\hat{\beta_c}\mathcal{C}_\ell(\psi).
\end{equation}
This indicates that the cutoff frequency is not independent; it is linked to both the onset and local resonance frequencies.
Lastly, the three types of characteristic frequencies obtained through beam theory are inserted into an element of the eigenmatrix $\mathbf{\hat{\Lambda}}$. 
Then, through a simple inverse matrix calculation from Eq.~(\ref{eq:S11}), we obtain the spring-mass parameters of a discretized unit cell.
\vspace{3mm}
\\
\indent
In Fig.~\ref{figS4}(a), we show the mass distribution of the discretized unit cell (\textit{single} cylindrical element), beginning with the first mode ($q=1$) at the bottom and increasing in mode numbers from the bottom to the top. 
The horizontal axis is the shift $\mu$ (i.e., contact location), and brighter color represents a larger mass distribution. 
The local mass in each mode is normalized to its maximum value.
Figures~\ref{figS4}(b)-(c) summarize the mass and stiffness of the first seven local resonators ($q=1$ to $7$) as a function of shift $\mu$.
It should be noted that while the distribution of mass and stiffness varies with the shift of the cylinder, the resulting local resonance frequency (i.e., $\omega_q=\sqrt{k_q/m_q}$) remains constant for each mode. This is because the cylinder's geometry and material properties do not change.
\\
\indent
We find that the mass and stiffness of the local resonator vary significantly with the shift $\mu$.
A key observation is that for anti-symmetric modes ($q$=~2,~4,~6,...), both the local mass and stiffness approach zero at the CoM ($\mu$=0).
As discussed in the main text, this suggests that the BICs from the anti-symmetric modes are supported by near-zero mass and stiffness.
Moreover, for both symmetric and anti-symmetric modes, we observe accidental-type BICs, where both local mass and stiffness approach zero at specific shift $\mu$.
All these BIC points arise from polarization-protected BICs, originating from the orthogonality between longitudinal waves from the contact boundary and the single resonator’s transverse waves.
\\
\indent
Mathematically, these near-zero mechanical coefficients can be identified using our DEM modeling approach.
In the dimensionless eigenmatrix [Eq.~(\ref{eq:S12})], the cutoff frequency and onset frequency are exactly the same, making the eigenmatrix singular. 
Consequently, the pseudo-inverse matrix yields a mechanical coefficient value of zero.
Physically, this equality of the cutoff and onset frequencies for the unit cylinder means that no energy transmission occurs in that mode. 
In other words, if energy is excited within the resonator, it remains completely localized inside the resonator.
\begin{figure}[htp]
\includegraphics [width=0.98\textwidth]{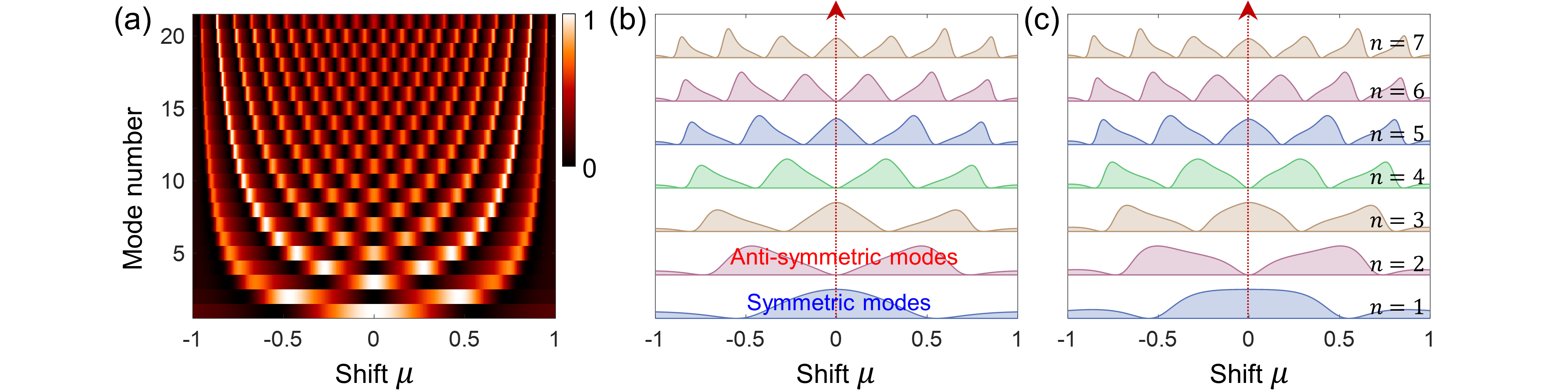}
    \caption{(a) Mass distribution of the discretized unit cell as a function of shift $\mu$ and mode number $q$. The local mass of each mode is normalized to its maximum value at a specific shift $\mu$. (b)-(c) Distribution of mass and stiffness of the first seven local resonators, where odd-numbered modes (i.e., $q$=~1,~3,~5, ...) correspond to the symmetric bending modes, and even-numbered modes (i.e., $q$=~2,~4,~6, ...) correspond to the anti-symmetric bending modes.}
\label{figS4}
\end{figure}
\newpage
\section{III.~Scattering Analysis\label{sec:sca}}
\begin{figure}[htp]
\includegraphics [width=0.98\textwidth]{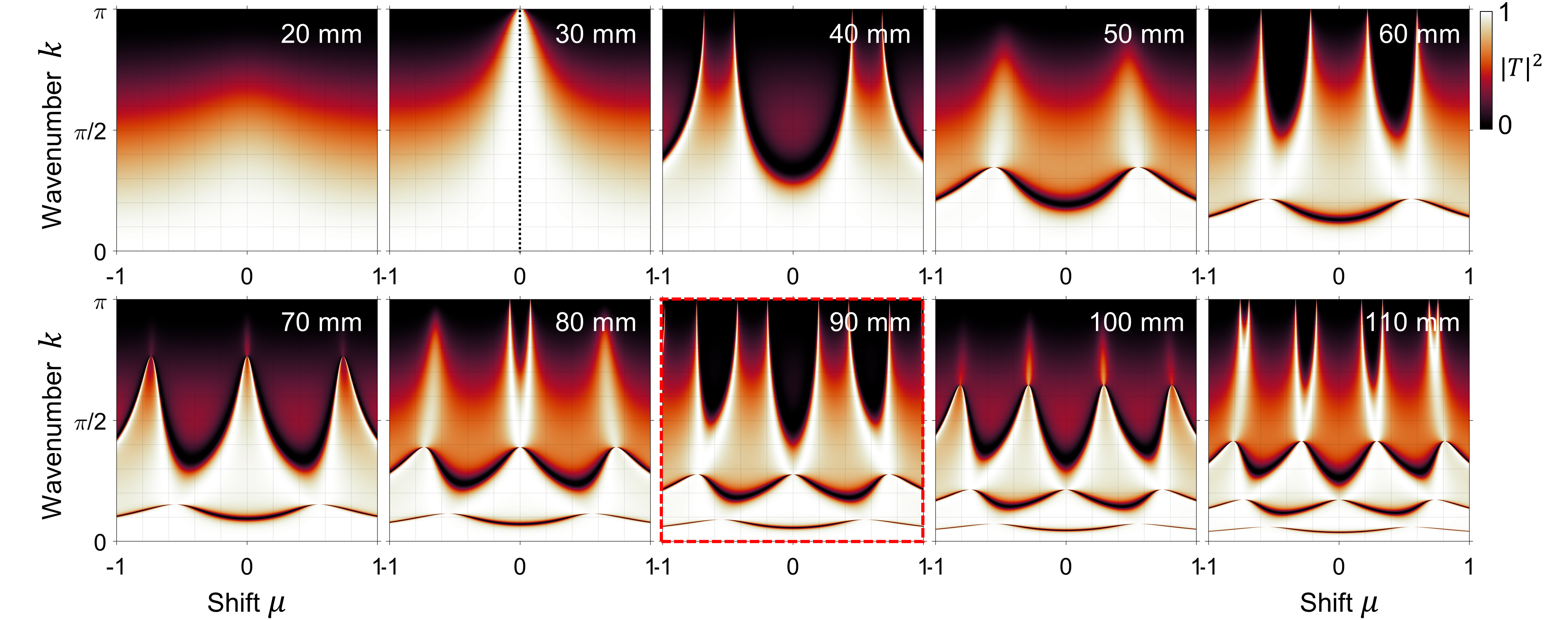}
    \caption{Transmission coefficients as a function of wavenumber $k$ and shift $\mu$ for various impurity lengths, with a host cylinder length of 30 mm.}
\label{figS5}
\end{figure}

Building on the discretized building blocks, we investigate the scattering processes between a plane wave $e^{i(kn-\omega t)}$ and a single defect in the linear regime.
By directly substituting Eq.~(\ref{eq:S3}) into Eq.~(\ref{eq:S2a}), we obtain the equation of motion for the $n$-th unit cell in a different form, as follows:
\begin{equation}\label{eq:S24}
    {{m}_{n}}\ddot{u}_{n}=\beta_c({u}_{n-1}-2{u}_{n}+{u}_{n+1})
    +\sum_{q}k_{n,q}\Delta_{n,q}u_n,
\end{equation}
where
\begin{equation}\label{eq:S25}
    \Delta_{n,q} = \frac{1}{{\omega_{n,q}^2}/{\omega^2}-1},
\end{equation}
We treat long cylinders (referred to as BIC defects in our study) as impurities perturbing a host of short cylinders as shown in Fig.~\textcolor{blue}{1}(a) of the main text. 
We explore the influence of the shifts of a single impurity on the resonance coupling strength.
To investigate the impact of impurities on the scattering problem, we use the following ansatz~\cite{Miroshnichenko2010}:
\begin{equation}\label{eq:S26}
    u_n = 
    \begin{cases}
    e^{i(kn-\omega{t})}+Re^{-i(kn+\omega{t})}, & \text{if}~n\leq0,\\
    Te^{i(kn-\omega{t})}, & \text{if}~n>0.
    \end{cases}
\end{equation}
This expression represents an incident plane wave interacting with the impurity, leading to the generation of reflected $(R)$ and transmitted waves $(T)$ near the defect sites $(n=0)$.
In this way, we define a transmission coefficient $|T|^2$ and a reflection coefficient $|R|^2$.
It should be noted that in linear dynamics, the conservation of energy and momentum does not allow for changes in the wavenumber or frequency of incoming and outgoing waves during the scattering process.
By substituting Eq.~(\ref{eq:S26}) into Eq.~(\ref{eq:S24}) near the impurities and carrying out a simple calculation, we obtain the following linear system of equations for $T$ and $R$
\begin{equation}
\mathcal{S}\begin{pmatrix}
  T \\
  R \\
\end{pmatrix}=\mathcal{R},
\end{equation}
with
\begin{equation}
\begin{aligned}
    \mathcal{S}&=\begin{pmatrix} \beta_{c}e^{ik} & \widetilde{m}_{0}\omega^2+\beta_{c}(e^{ik}-2)+\sum_{q}\widetilde{k}_{q}\widetilde{\Delta}_{q} \\ {m}_{0}\omega^{2}e^{ik}+\beta_{c}e^{ik}(e^{ik}-2)+\sum_{q}k_{q}\Delta_{q}e^{ik} & \beta_{c} \end{pmatrix},\\
    \mathcal{R}&=\begin{pmatrix}
     -\widetilde{m}_0\omega^2+\beta_{c}(2-e^{-ik})-\sum_{q}\widetilde{k}_{q}\widetilde{\Delta}_{q} \\
     -\beta_c \\
    \end{pmatrix}.
\end{aligned}
\end{equation}
Here, the tilde notation represents indexing related to impurities.
Inherently, $|T|^2$ and $|R|^2$ are complementary quantities; an increase in the other always accompanies a reduction in one, and vice versa.
Due to the short length of the host cylinder, $\omega_q\gg \omega$ in the low-frequency range, leading to ${\Delta}_q\approx{0}$. 
As a result, this configuration supports a broad acoustic band akin to a classical monatomic chain.
\\
\indent
In Fig.~\ref{figS5}, we show the transmission coefficients as a function of wavenumber $k$ and the shift $\mu$ for various impurity lengths.
Firstly, we observe an accidental-type BIC in the symmetric mode when the BIC defect length is 50 mm. 
Upon further increasing the defect length to 70 mm, a symmetry-protected BIC emerges in the anti-symmetric mode. 
These mode symmetries are verified by the number of BIC points. 
As the length of the BIC defect increases, multiple BIC points continue to appear. 
This occurs because, as the cylinder length increases, the bending mode frequency follows a relationship with $\omega\sim{1/L^2}$~\cite{inman1994}.
\newpage
\section{IV.~Steady state-space\label{sec:sss}}
In this section, we discuss the frequency response for calculating the Q-factor in the finite chain.
Considering the effects of energy dissipation in the system, the equation of motion for a chain of $N$ particles can be expressed as follows:
\begin{subequations}\label{eq:S29}
\begin{eqnarray}
m_{1,0}\frac{d^2u_1}{dt^2}&=&-\beta_{w1} u_1 - \beta_c(u_1-u_2)+\sum_{q}k_{1,q}(u_{1,q}-u_1)-\gamma(v_1-v_2), \label{eq:S29a}\\
m_{n,0}\frac{d^2u_n}{dt^2}&=&{\beta}_c\left( u_{n-1} - 2 u_{n} + u_{n+1} \right) +\sum_{q}k_{n,q}(u_{n,q}-u_{n})+\gamma\left( v_{n-1} - 2 v_{n} + v_{n+1} \right), \label{eq:S29b}\\
m_{N,0}\frac{d^2u_N}{dt^2}&=&-\beta_{w2} u_N+\beta_c(u_{N-1}-u_N) +\sum_{q}k_{N,q}(u_{N,q}-u_N)+\gamma(v_{N-1}-v_{N}),\label{eq:S29c}
\end{eqnarray}
\end{subequations}
where $\beta_{w1(2)}$ is the linearized contact stiffness between the first (last) particle and the actuator (sensor); $v_i$ is the particle velocity.
Among the various dissipation models, we adopt a linear viscous damping model between neighboring particles. 
This is because, in our experimental setup, the energy loss from contact behavior (i.e., solid friction and resulting thermal effects) is the most prominent.
By combining the equation of motion in Eq.~(\ref{eq:S29}) with the equation of motion for the local resonator, we can construct the dynamic system in state-space form as follows:
\begin{equation}
    \begin{aligned}
        \dot{\bm{X}}&=\bm{A}\bm{X}+\bm{B}\text{F}_d,\\
        \text{F}_N&=\bm{C}\bm{X}+\bm{D}\text{F}_d.
    \end{aligned}
\end{equation}
Here, $\bm{X}$ is the state vector, and $\bm{A},~\bm{B},~\bm{C},$ and $\bm{D}$ represent the system, input, output, and direct transmission matrices, respectively.
Because there is no direct input affecting the output, $\bm{D}$ should be a null state.
$F_d$ indicates the system's input corresponding to the dynamic force applied to the first particle, and $F_{N}$ represents the dynamic force transmitted at the system's end.
$\bm{X}$,~$\bm{A}$,~$\bm{B}$, and $\bm{C}$ are given by:
\begin{equation}
\bm{u} = \begin{bmatrix}
u_1 \\
\vspace{-4pt} \vdots \\
u_N \\
u_{1,1} \\
\vspace{-4pt} \vdots \\
u_{N,1} \\
\vspace{-4pt} \vdots \\
u_{1,q} \\
\vspace{-4pt} \vdots \\
u_{N,q} \\
\end{bmatrix},\quad
\bm{X} = \begin{bmatrix}
\bm{u} \\
- \\
\dot{\bm{u}}
\end{bmatrix},\quad
\bm{A}= 
\begin{bmatrix}
0 & \bm{I} \\
\bm{M}^{-1}\bm{K} & \bm{M}^{-1}\bm{\Gamma} 
\end{bmatrix},\quad
\bm{B}=\begin{bmatrix}
\vdots \\
0 \\
\vdots \\
- \\
\frac{1}{m_{1}} \\
\vdots \\
0 \\
\vdots \\
\end{bmatrix},\quad
\bm{C}=\begin{bmatrix}
0 \\
\vspace{-3pt} \vdots \\
\beta_{w2} \\
\vspace{-3pt} \vdots \\
0 \\
- \\
\vspace{-3pt} \vdots \\
0 \\
\vspace{-3pt} \vdots \\
\end{bmatrix}^{T},
\end{equation}
Here, the bar in the column vector indicates the middle position.
$\bm{M}$, $\bm{\Gamma}$, and $\bm{K}$ denote the mass, damping, and stiffness matrices derived from the equation of motion of finite systems, respectively; $\bm{I}$ is the identity matrix.
For simplicity, we assume that the two boundary stiffness values, $\beta_{w1}$ and $\beta_{w2}$, are identical to $\beta_c$.
\\
\indent
We obtain the transfer function of the system and calculate its transmission gain using MATLAB's Bode function.
Within the frequency response, we determined the Q-factor of each resonant mode through the full width at half maximum (FWHM) in the frequency domain.
\newpage
\section{V.~Numerical Results\label{sec:num}}
In this section, we corroborate the results using the finite element method (FEM, COMSOL Multiphysics).
By conducting a full 3D solid mechanics analysis, we investigate the displacement field of resonators as they transition from BICs to quasi-BICs.
\begin{figure}[htp]
\includegraphics [width=0.9\textwidth]{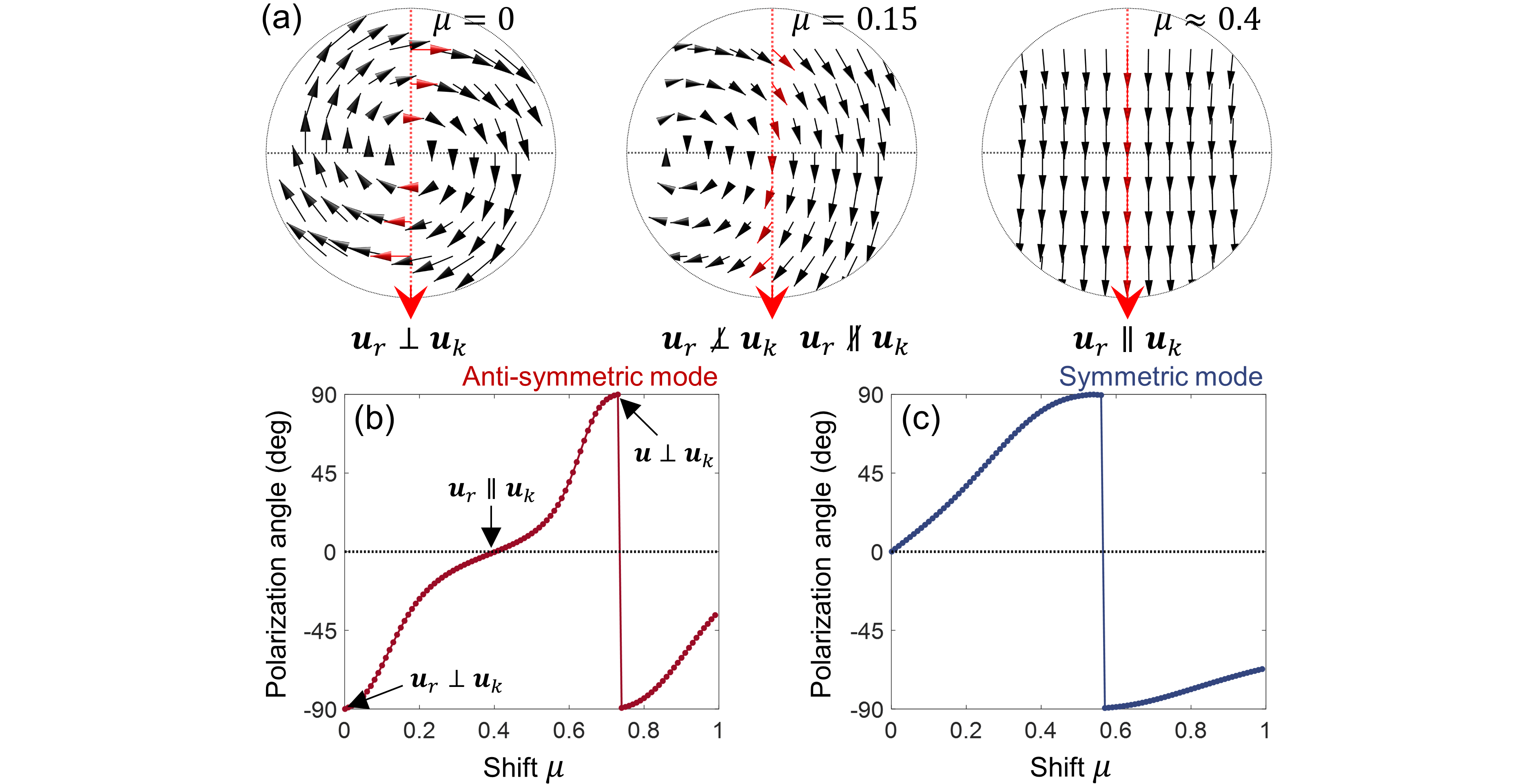}
    \caption{(a) Three types of displacement vector fields on the lateral surface of the resonators, depending on the polarization alignment between $\mathbf{u}_r$ (the displacement vector of the resonator's internal fields) and $\mathbf{u}_k$ (the displacement along the contact axis). The red arrow indicates the contact boundary axis. (b) The polarization angle for both anti-symmetric and symmetric modes. An orthogonal polarization angle (i.e., $\mathbf{u}_r$$\bot$$\mathbf{u}_k$) indicates polarization-protected BICs, while an angle of zero (i.e., $\mathbf{u}_r$$\parallel$$\mathbf{u}_k$) indicates a strong coupling effect.}
\label{figS6}
\end{figure}
\\
\indent
In Fig.~\ref{figS6}(a), we present the three types of displacement vector fields on the lateral surface [refer to Fig. 2(b) in the main text] within the resonators, with the red arrow indicating the contact boundary axis.
The left panel illustrates the polarization-protected BIC (i.e., $\mathbf{u}_r$$\bot$$\mathbf{u}_k$), while a further increase in the shift $\mu$ leads to a strong coupling between wave propagation and the internal resonance of the resonators (i.e., $\mathbf{u}_r$$\parallel$$\mathbf{u}_k$).
Through $\mathbf{u}_r$ and $\mathbf{u}_k$, we calculate the polarization angle for both anti-symmetric and symmetric modes, as shown in Fig.~\ref{figS6}(b).
When the polarization angle is orthogonal (i.e., $\pm{90}$ degrees), it indicates polarization-protected BICs, whereas an angle of zero indicates strong coupling effects.
The full evolution of the displacement fields on the lateral and contact surfaces is summarized in Figs.~\ref{figS7} and~\ref{figS8}, respectively.
These results align well with our theoretical DEM model and the corresponding wave dynamics.
\begin{figure}[htp]
\includegraphics [width=0.98\textwidth]{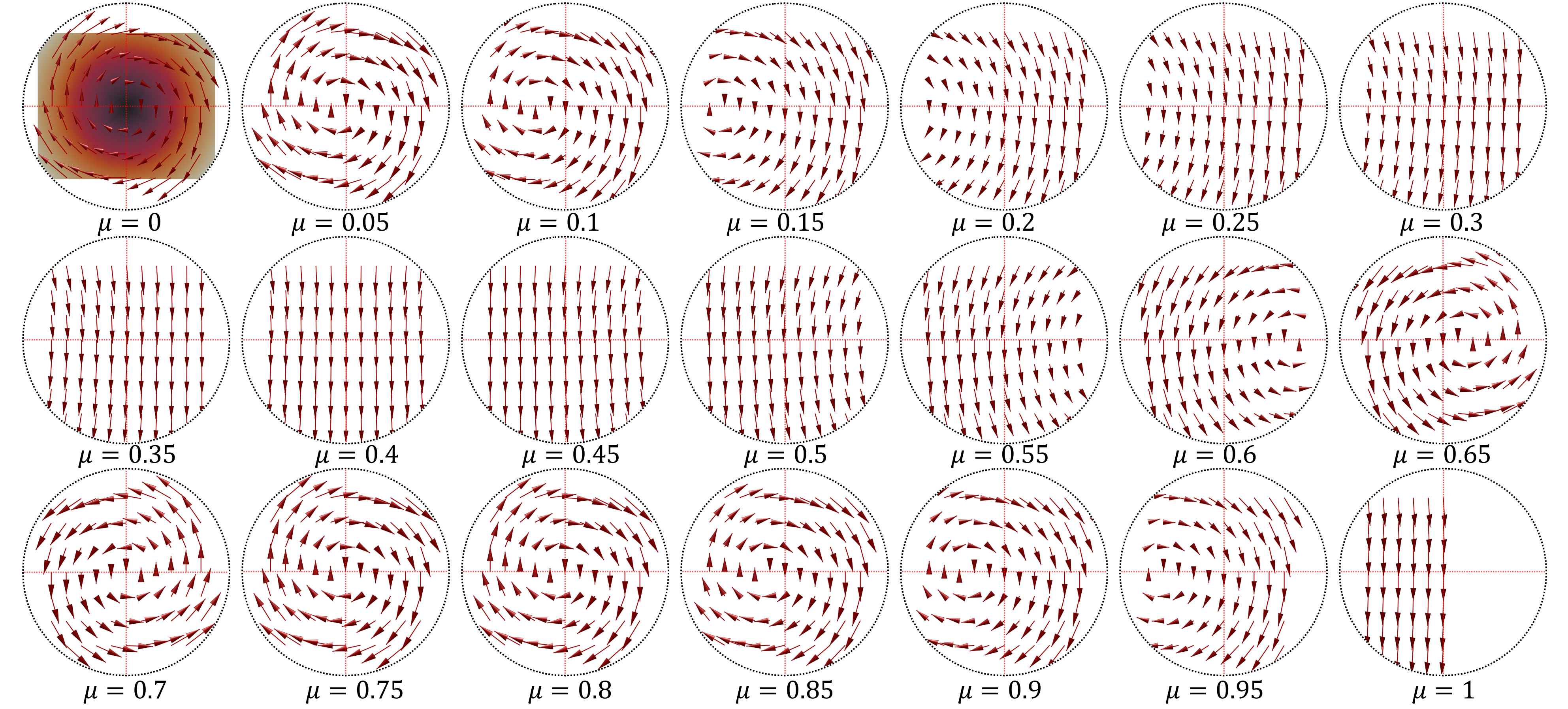}
    \caption{The evolution of the displacement vector fields on the lateral surface [see the main text of Fig.~2(b)] of the BIC defect cylindrical resonator as the shift $\mu$ increases.}
\label{figS7}
\end{figure}
\begin{figure}[htp]
\includegraphics [width=0.98\textwidth]{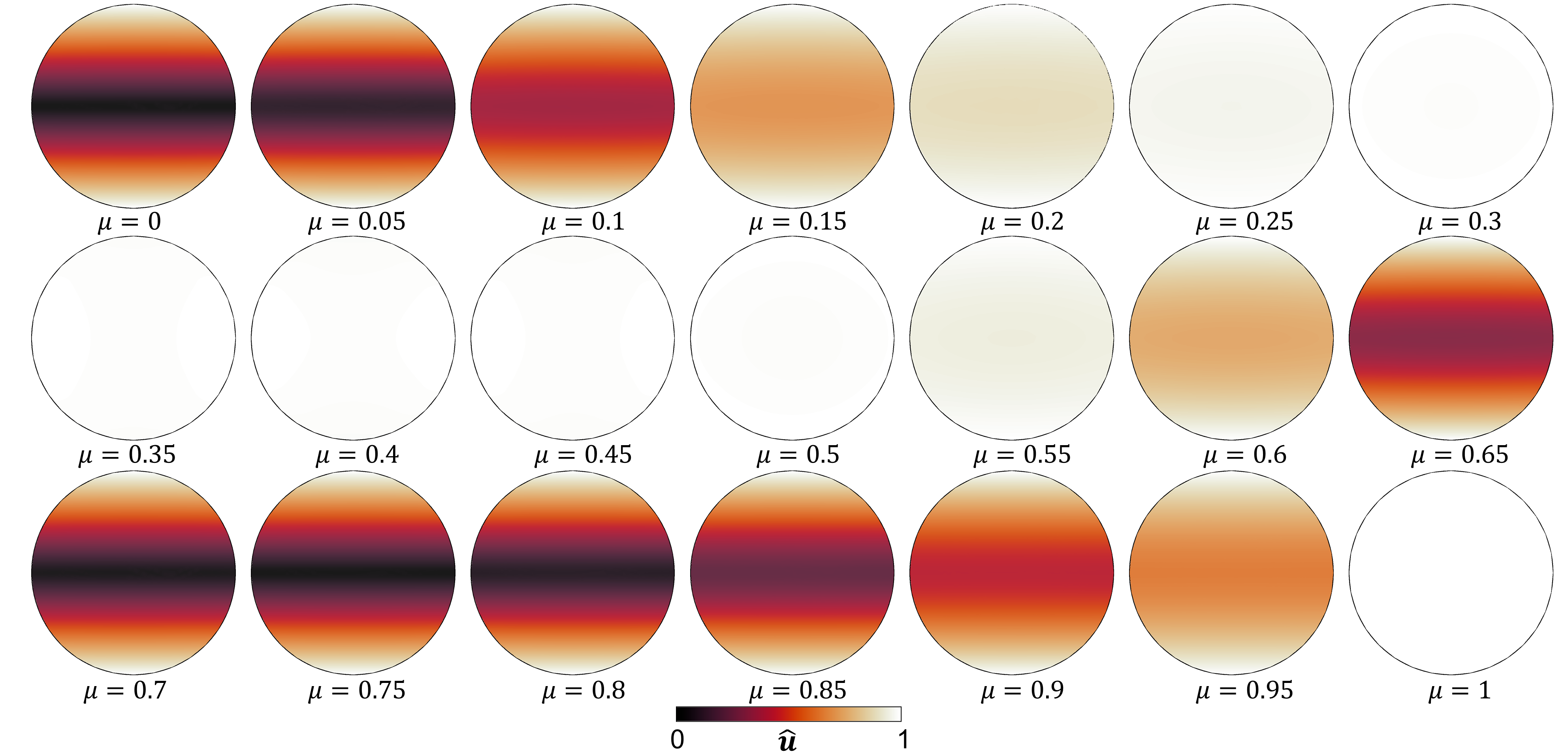}
    \caption{The evolution of the displacement intensity map on the contact surface [see the main text of Fig.~2(b)] of the BIC defect cylindrical resonator as the shift $\mu$ increases. The color indicates the normalized displacement.}
\label{figS8}
\end{figure}
\\
\newpage
\section{VI.~Asymmetry \& Damping effects of Q-factor\label{sec:eff}}
\begin{figure}[htp]
\includegraphics [width=0.99\textwidth]{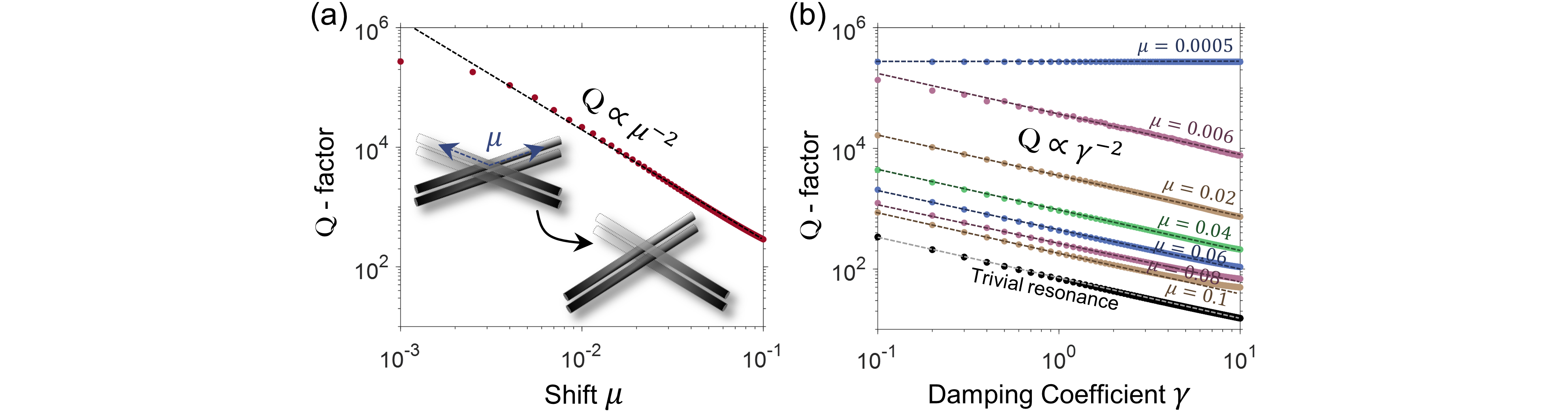}
    \caption{(a) Influence of shift $\mu$ (i.e., asymmetry of the resonators) on the Q-factor of the BIC. The Q-factor follows the $Q\propto\mu^{-2}$. (b) Influence of the damping coefficient $\gamma$ on the Q-factor of the BIC, considering a linear viscous damping model between the unit cells. Each of the multiple lines corresponds to calculations at different shifts $\mu$ for the bound mode, while the lowest black line represents the result for the trivial structural resonance.}
\label{figS9}
\end{figure}

In this section, we investigate the influence of two factors on the Q-factor of BICs: asymmetry of geometry and the damping coefficient.
The influence of asymmetry on the Q-factor is known to follow the law of $Q\propto\alpha^{-2}$, as established by previous work~\cite{Koshelev2018}.
Similarly, our system demonstrates that the Q-factor follows $Q\propto\mu^{-2}$ with respect to the shift $\mu$ [see Fig.~\ref{figS9}(a)].
\\
\indent
To understand the influence of damping effects, it is necessary to consider non-dissipative systems.
Theoretically, in systems without any dissipation effects, the Q-factor of all resonances, whether BICs or structural resonances, diverges.
However, by introducing damping effects, the difference between these two types of resonances becomes clearer.
Among the various energy dissipation effects in the system, we consider a linear viscous damping model between the unit cells, as the primary source of energy dissipation in our granule system arises from contact behavior, such as friction and thermal effects.
In Fig.~\ref{figS9}(b), we present the Q-factor profiles as a function of the linear damping coefficient $\gamma$.
Each of the multiple lines corresponds to calculations at different shifts $\mu$; the lowest black line represents the result for the trivial structural resonance.
We observe that the damping effect exhibits a similar trend to the $\alpha^{-2}$ law associated with asymmetry.
An exception occurs when $\mu=0.0005$, which corresponds to a very small shift near the exact BIC points.
Interestingly, even as the damping coefficient increases (at least up to the considered $\gamma$ values), there is no significant effect on the Q-factor.
This occurs because the mechanical coefficient is nearly zero, resulting in zero mode coupling, and consequently, the coupling to damping is also negligible.
We further observe that the Q-factor of quasi-BICs is consistently higher than that of trivial structural resonances, attributed to the unique mechanisms inherent to BICs.
\newpage
\section{VII.~Experimental Support of Bound Bands in the Continuum (BBICs)\label{sec:bbic}}
\subsection{A.~Time evolution}
\begin{figure}[htp]
\includegraphics [width=0.99\textwidth]{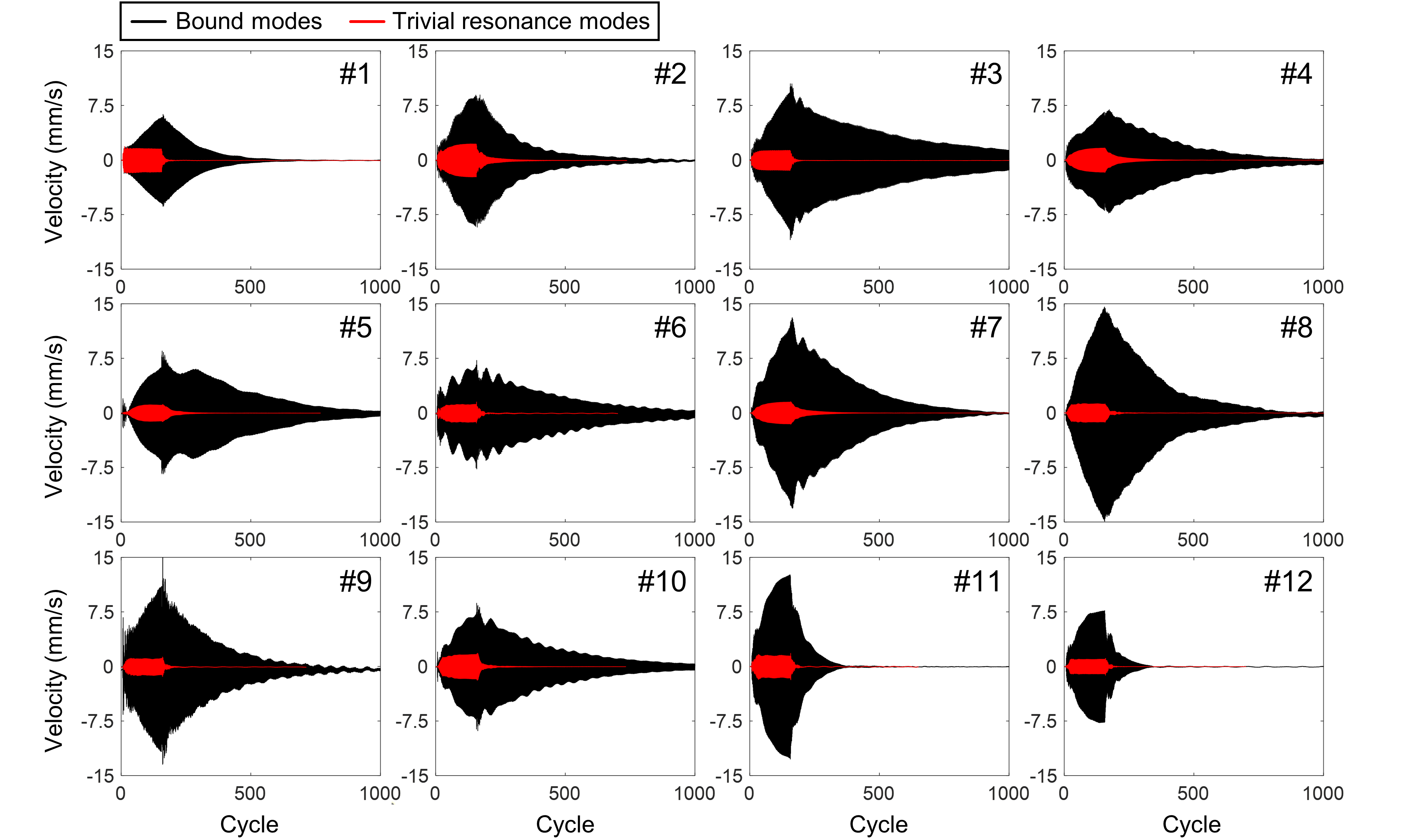}
    \caption{(a) Experimentally measured velocity profiles for all cylindrical resonators in the chain under quasi-BBIC conditions. The system is stimulated for 150 cycles at the frequency corresponding to the bound modes or structural resonance modes. The black line represents the bound modes, while the red lines correspond to the trivial structural resonances. 
    The symbol \# denotes the particle index; specifically, the bottom of the chain is \#1, while the top of the chain is \#12.
    For the bound modes, all cylindrical resonators in the chain show a very slow energy radiation after 150 cycles, indicating a high Q-factor.}
\label{figS10}
\end{figure}

In this section, we provide experimental support for the BBICs presented in the main text.
Experimentally, the cylindrical chain is initially aligned orthogonally at the center of mass (CoM) contact points, using cylinders with a length of 110 mm. 
Following this, we introduce a slight shift to each resonator to break their symmetry. The chain is then excited locally at its initial segment, using the bound mode frequency over 150 cycles. 
Localized velocity profiles of vibrations near the contact points of each cylindrical resonator are measured using a laser Doppler vibrometer.
\\
\indent
In Fig.~\ref{figS10}, we present the velocity profiles for all cylindrical resonators, where the black line represents the bound modes and the red lines correspond to the trivial structural resonances.
The structural resonances arise from multiple peaks within the passband of the finite chain, and we have selected the frequency that is closest to the bound mode.
We clearly observe that the bound modes consistently accumulate energy with the excited frequency, while the trivial resonances also build up energy but quickly reach saturation.
Once the excitation ceases (after 150 cycles), the localized energy associated with the bound modes remains in the cylindrical resonators for a long time, which can be explained by the high Q-factor of BICs.
On the other hand, the trivial resonance modes show rapid energy dissipation, similar to a damped harmonic oscillator with a high damping coefficient.
In the chain's boundaries, specifically at the initial and end parts of the chain, we observe lower energy accumulation compared to the chain's interior, which can be attributed to boundary effects in the experiment.
\subsection{B.~Comparing Q-factor}
\begin{figure}[htp]
\includegraphics [width=0.99\textwidth]{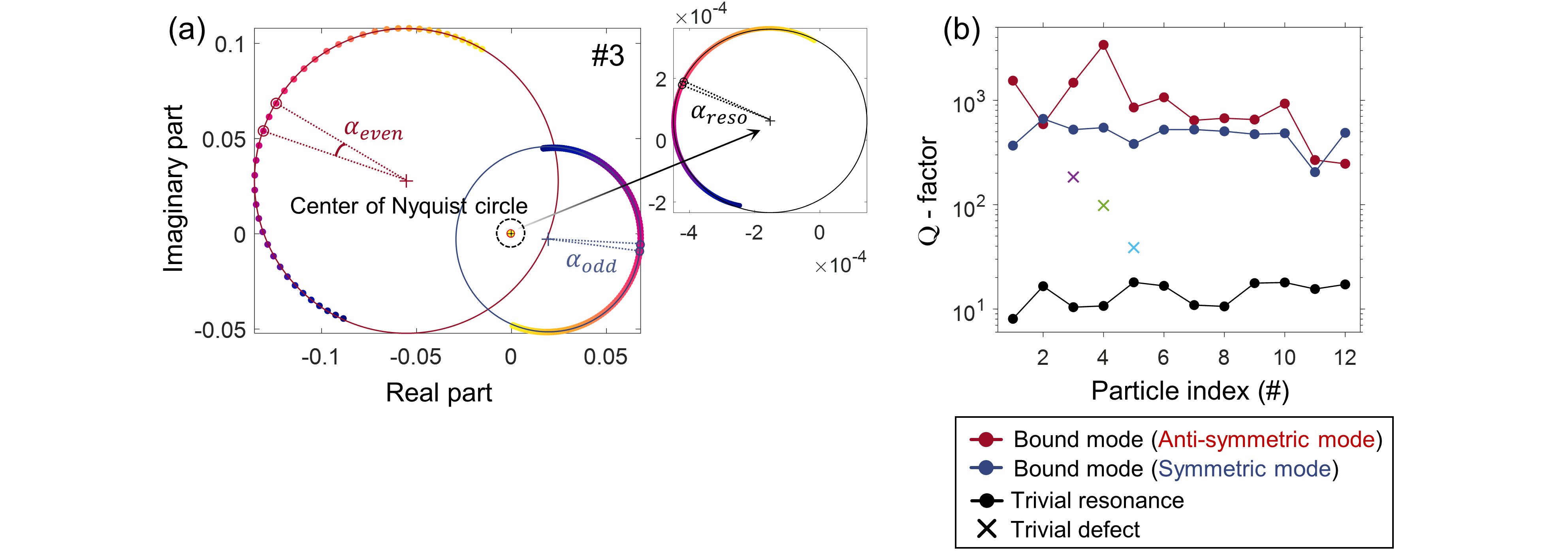}
    \caption{(a) Experimentally obtained Nyquist plot for the third cylindrical resonator within the chains, featuring three circles corresponding to the excitation modes: the largest circle arises from the anti-symmetric bound mode, the second largest from the symmetric bound mode, and the smallest from the trivial resonance mode. (b) Calculated Q-factors for the entire chain.}
\label{figS11}
\end{figure}

We calculate the Q-factor for each cylindrical resonator here.
To achieve a more precise Q-factor, we employ the Nyquist plot method~\cite{inman1994}, which is considered more accurate than the FWHM approach.
In damped structures, as the frequency increases, the trajectory of the real and imaginary components of the frequency response takes on a circular shape around the natural frequency. 
In Fig.~\ref{figS11}(a), we present the Nyquist plot for the third cylindrical resonator within the chains. 
The points on the circle represent the displacement response in the complex plane, while the solid line is an approximate circle fit.
Figure~\ref{figS11}(a) includes three circles: the largest one arises from the anti-symmetric bound mode, the second largest from the symmetric bound mode, and the smallest from the trivial resonance mode.
A smaller circle in the Nyquist plot indicates larger damping at similar frequencies.
The damping derived from the Nyquist plot can be calculated using the following formula:
\begin{equation}
    \zeta = \frac{\omega_b^2-\omega_a^2}{2\omega_n[\omega_a\tan({\alpha/2})+\omega_b\tan({\alpha/2})]}.
\end{equation}
where $\omega_n$ is the damped natural frequency, and $\omega_a$ and $\omega_b$ are the two frequencies that are 3 dB lower than the peak. $\alpha$ represents the angle between $\omega_a$ and $\omega_b$ on the Nyquist circle [refer to Fig.~\ref{figS11}(a)].
Using this, we calculate the Q-factor with the following relationship
\begin{equation}
    Q=\frac{1}{2\zeta}.
\end{equation}
In Fig.~\ref{figS11}(b), we summarize the Q-factors of these modes according to the particle index.
We observe that high Q-factors are maintained within the cylinders for both symmetric and anti-symmetric bound modes.
This characteristic arises because each single cylindrical resonator supports BICs.
In contrast, as shown by the black points, the trivial resonance mode exhibits a significantly lower Q-factor.
Additionally, we investigated the effect of energy localization due to trivial defects (i.e., resonance modes resulting from defects localized within the bandgap of the cylindrical chain).
We measured the Q-factors of the corresponding defect modes by shifting the position of the trivial defect from \#3 to \#5.
Our results indicate that as the defect moves further into the interior of the chain, the wave localization characteristics associated with it weaken significantly due to the nature of evanescent waves.

